\documentclass[journal=jctc,manuscript=article]{achemso}

\usepackage[utf8]{inputenc}
\usepackage[T1]{fontenc}
\usepackage{mathptmx}
\usepackage{wrapfig}                    
\usepackage{graphicx}				    
\usepackage{dcolumn}                    
\usepackage{amssymb, amstext, amsmath}  
\usepackage{bm}                         
\usepackage{nicefrac}                   
\usepackage{tabularx,booktabs}          
\usepackage{soul,xcolor}                
\usepackage{verbatim}                   
\usepackage{float,placeins}             
\usepackage{dsfont}
\usepackage{blkarray}
\usepackage{setspace}                   

\usepackage{accents}
 \newcommand*{\dt}[1]{%
  \accentset{\mbox{\large\bfseries .}}{#1}}
\newcommand*{\ddt}[1]{%
  \accentset{\mbox{\large\bfseries .\hspace{0.0ex}.}}{#1}}
  
\SectionNumbersOn

\title[\\]{A theory of phonon-induced friction on molecular adsorbates}

\author{Ardavan Farahvash}
\affiliation{Department of Chemistry, Massachusetts Institute of Technology, Cambridge, Massachusetts 02139, USA}

\author{Adam P. Willard}
\email{awillard@mit.edu}
\affiliation{Department of Chemistry, Massachusetts Institute of Technology, Cambridge, Massachusetts 02139, USA}

\begin{document}

\singlespacing

\begin{abstract}
    In this manuscript, we provide a general theory for how surface phonons couple to molecular adsorbates. Our theory maps the extended dynamics of a surface's atomic vibrational motions to a generalized Langevin equation, and by doing so captures these dynamics in a single quantity: the non-Markovian friction. The different frequency components of this friction are the phonon modes of the surface slab weighted by their coupling to the adsorbate degrees of freedom. Using this formalism, we demonstrate that physisorbed species couple primarily to acoustic phonons while chemisorbed species couple to dispersionless local vibrations. We subsequently derive equations for phonon-adjusted reaction rates using transition state theory and demonstrate that these corrections improve agreement with experimental results for CO desorption rates from Pt(111).
\end{abstract}

\maketitle

\section{Introduction}
    \label{sec:intro}
    The atomic vibrations of a solid surface can significantly influence the rates and mechanisms of surface chemical processes.\cite{chen_computational_2021,norskov_density_2011,koch_adsorption-induced_2008,michalsky_departures_2014}
    Unraveling these influences experimentally is difficult, necessitating the development of theoretical tools. In this manuscript, we introduce a theory for the coupling between the nuclear dynamics of surface-bound (adsorbed) molecules and the vibrations of the underlying surface.
    Our theory projects the collective surface vibrations onto the motion of the adsorbate, and in doing so captures how surface phonons spanning several order of magnitude in length/energy scale influence adsorbate dynamics. 

    Over the past four decades, a multitude of experimental and theoretical studies have demonstrated how surface vibrations affect reaction dynamics at solid interfaces.
    For example, several studies have shown that surface atom motion plays a central role in the dissociation of methane and molecular nitrogen on a variety of different metal surfaces. \cite{luntz_activation_1989,luntz_ch4_1991,henkelman_theoretical_2001,nave_methane_2007,nave_methane_2009,nave_methane_2010,tiwari_methane_2009,tiwari_temperature_2010,egeberg_molecular_2001,diekhoner_n2_2001,nattino_n2_2015,shakouri_accurate_2017,shakouri_analysis_2018}. 
    Certain experiments have utilized piezoelectric drivers to generate surface acoustic waves and demonstrated that waves of particular frequencies and polarizations are capable of enhancing some reactions, such as the oxidation of ethanol or carbon monoxide.\cite{mitrelias_effect_1998,kelling_photoemission_1998,kelling_surface_1999,nishiyama_effects_2000,nishiyama_iras_2005,nishiyama_peem_2006,inoue_effects_2007,von_boehn_promotion_2020}. 
    While several mechanisms have been proposed to explain this rate enhancement, the origin of this effect remains largely uncertain.\cite{von_boehn_promotion_2020}
    Much of this uncertainty stems from the inability of current theory to connect the lengthscales and timescales of atomistic dynamics with the mesoscopic scale of experimentally relevant surface acoustic waves \cite{an_quantum_2016}. 
    The theoretical approach that we present here is capable of effectively connecting these scales and thereby providing new physical insight into the roles of surface vibrations in surface chemical reactions.
    
\begin{figure}[tbhp]
        \centering
        \includegraphics*[width=3.25in]{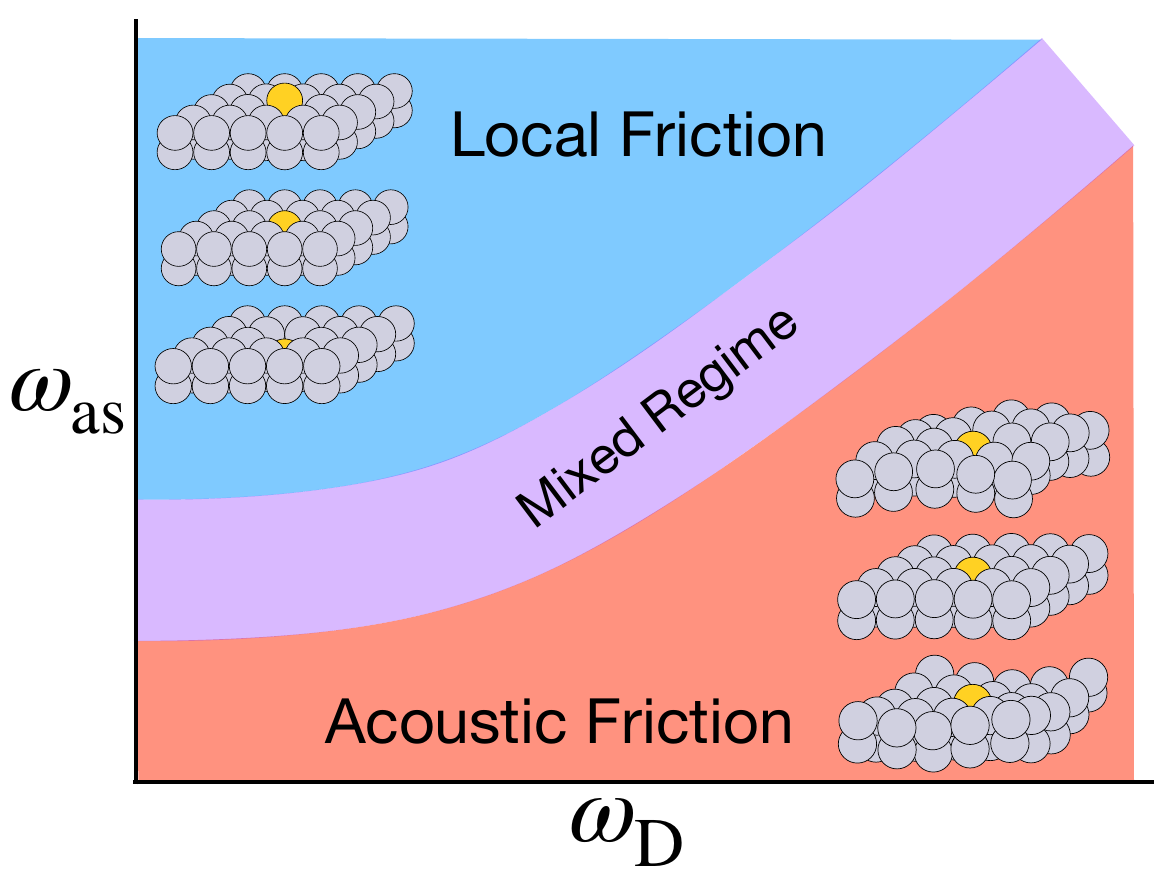}
       \caption{ 2D schematic illustrating the dominant phonon modes in terms of their coupling to the adsorbate or contribution to the friction kernel. $\omega_{\mathrm{as}}$ is the frequency of the adsorbate-surface bond and the $\omega_{\mathrm{D}}$ is the solid's Debye frequency.}
        \label{figure_intro}
    \end{figure}

    Our theoretical model describes the motion of surface-bound molecules via a generalized Langevin equation (GLE). The central parameter of this equation is the non-Markovian friction kernel (or memory kernel), which encapsulates the different surface vibrational modes weighted by their coupling to the adsorbate. 
    This friction kernel can be developed to represent surface modes with wavelengths well beyond that of a typical simulation cell, thereby eliminating a common source of finite-size effects.
    Using this formalism, we demonstrate that the influence of surface phonons on the dynamics of an adsorbed molecule depends significantly on the magnitude of the adsorbate-surface coupling.
    Specifically, we find that chemisorbed species couple primarily to dispersionless local vibrations, while physisorbed species couple primarily to acoustic phonons across a broad range of frequencies.
    The key parameter that determines whether the primary influence of surface vibrations is via extended or localized phonon modes is the ratio of the solid's Debye frequency to the frequency of the adsorbate-surface bond (Fig.~\ref{figure_intro}).
    By combining these observations with harmonic transition state theory, we derive equations that describe how phonons alter reaction rates at solid surfaces, and demonstrate that these phononic corrections agree with experimental measurements of desorption rate constants.

    The approach we present in this manuscript extends a previous method for utilizing the GLE for describing the influence of surface vibrational modes on molecular adsorbates.~\cite{tully_dynamics_1980,farahvash_modeling_2023} 
    In the previous method, the GLE was designed to describe the collective influences of the surrounding solid on a single representative surface atom (\textit{i.e.}, binding site). 
    In the method we present here, the GLE is designed to describe the influence of the entire surface on the dynamics of the adsorbate itself.
    This new approach is more easily interpretable because the friction kernel directly represents influence of surface phonons on the adsorbate.
        
    The remainder of the paper is organized as follows. 
    In Section \ref{sec:phonon_induced_friction}, we present the formal theory behind the phonon-induced GLE and highlight the characteristics of the friction kernel. Along with the derivation of our theoretical method, we present its application to a simple model system of an adsorbate whose surface bond has a tunable frequency. We then analyze results for the friction kernel across a range of values for adsorbate-surface bond frequency. 
    In Section~\ref{sec:dispersion}, we provide further insight into the results presented in Section~\ref{sec:phonon_induced_friction} by analyzing the role of dispersion plays in the phonon-induced friction.
    Finally, in Section~\ref{sec:rates}, we develop a formulation of the transition state theory desorption rate that explicitly accounts for the influence of surface phonon modes. We validate this formulation by demonstrating an improved agreement with experiments of CO desorption from Pt(111).

\section{Coupling surface vibrations to adsorbates}
\label{sec:phonon_induced_friction}

    \subsection{Theory}
    \label{subsec:sec2_theory}

    In this section, we derive a theory for modeling the effects of surface vibrations on the dynamics of an adsorbate.
    The general approach involves integrating out the nuclear degrees of freedom of the solid and representing their influence as effective equations of motion for the adsorbate.
    In order to do so we must make two critical assumptions. First, we assume that the nuclear degrees of freedom of the adsorbate and solid can be described classically and adiabatically. There are cases where adsorbate dynamics are non-adiabatic, and in such cases a mixed quantum-classical GLE approach has been shown to be effective \cite{headgordon_molecular_1995,hertl_random_2021}, however the development and analysis of such a theory is beyond the scope of this paper.
    Second, we assume that the solid potential energy surface and surface-adsorbate interaction are both harmonic. 
    This approximation simplifies the formulation of the theory and is appropriate for the development of generalizable physical insight.
    We note that the effects of anharmonicity can be incorporated into the theory, which would be a natural target of future developments.
    
    To begin, we define $\mathbf{x}_\mathrm{A}$ and $\mathbf{x}_\mathrm{S}$ as the the mass-weighted displacements of the adsorbate and solid nuclei, respectively, from their equilibrium positions. We decompose the total potential energy into contributions from the adsorbate and solid, 
    \begin{equation}
        \label{eq:PES}
        V(\mathbf{x}_\mathrm{A},\mathbf{x}_\mathrm{S}) = V_\mathrm{A}(\mathbf{x}_\mathrm{A}) + V_\mathrm{AS}(\mathbf{x}_\mathrm{A},\mathbf{x}_\mathrm{S}) + V_{S}(\mathbf{x}_\mathrm{S}),\\
    \end{equation}
    where $V_\mathrm{A}$ is the adsorbate potential energy containing all intramolecular and intermolecular interactions, $V_\mathrm{S}$ is the solid potential energy surface, and $V_\mathrm{AS}$ the adsorbate-solid interaction. 
    A second-order expansion of $V_\mathrm{AS}$ and $V_\mathrm{S}$ around the equilibrium nuclear positions invites the definition of the matrices $\mathbf{G}_\mathrm{AS}$, $\mathbf{G}_\mathrm{A}$, $\mathbf{G}_\mathrm{S}$, and $\mathbf{H}_\mathrm{S}$, whose elements are given by,
    \begin{gather}
        \label{eq:G_AS}
        G_{ \mathrm{AS};ij} = \frac{\partial^2 V_\mathrm{AS}}{\partial {x}_{\mathrm{A};i}  \partial {x}_{\mathrm{S};j}},\\
        \label{eq:G_A}
        G_{ \mathrm{A} ;ij} = \frac{\partial^2 V_\mathrm{AS}}{\partial {x}^2_{\mathrm{A};i}} \delta_{ij}, \\
        \label{eq:G_S}
        G_{ \mathrm{S} ;ij} = \frac{\partial^2 V_\mathrm{AS}}{\partial {x}^2_{\mathrm{S};i}} \delta_{ij}, \\
        \label{eq:H_S}
        H_{ \mathrm{S};ij} = G_{\mathrm{S};ij} + \frac{\partial^2 V_\mathrm{S}}{\partial {x}_{\mathrm{S};i}  \partial {x}_{\mathrm{S};j}},
    \end{gather}
    where $\delta_{ij}$ is the Kronecker delta, $\mathbf{H}_\mathrm{S}$ is the solid's mass-weighted Hessian, $\mathbf{G}_\mathrm{AS}$ is the coupling between adsorbate and solid degrees of freedom, and $\mathbf{G}_\mathrm{A}$ and $\mathbf{G}_\mathrm{S}$ are diagonal matrices of oscillation frequencies for the adsorbate and solid degrees of freedom respectively. 
    All derivatives in Eqs.~\ref{eq:G_AS}-\ref{eq:H_S} are evaluated the equilibrium nuclear positions.
    
    The expansion of $V_\mathrm{AS}$ and $V_\mathrm{S}$ can be manipulated to generate equations of motion for adsorbate, 
    \begin{equation}
        \label{eq:eom_1A}
        \ddt{\mathbf{x}}_\mathrm{A} = -\frac{\partial V_\mathrm{A}}{\partial \mathbf{x}_\mathrm{A}} - \mathbf{G}_\mathrm{A} \mathbf{x}_\mathrm{A} - \mathbf{G}_\mathrm{AS} \mathbf{x}_\mathrm{S},\\
    \end{equation}
    and solid atoms,
    \begin{equation}
        \label{eq:eom_1B}
        \ddt{\mathbf{x}}_\mathrm{S} = -\mathbf{G}^T_\mathrm{AS} \mathbf{x}_A - \mathbf{H}_{S} \mathbf{x}_\mathrm{S}.
    \end{equation}
    By utilizing a second-order expansion, we impose a harmonic approximation on the model.
    It is convenient to describe the nuclear displacements of the solid in a basis of phonon modes.
    Thus, we introduce solid phonon coordinates, $\mathbf{u}_\mathrm{S} = \mathbf{U}^T \mathbf{x}_\mathrm{S} $, here $\mathbf{U}$ is a matrix with the eigenvectors of $\mathbf{H}_\mathrm{S}$ as columns. In terms of these coordinates the equations of motion may be expressed as,
    \begin{gather}
        \label{eq:eom_2A}
        \ddot{\mathbf{x}}_\mathrm{A} =  -\frac{\partial V_\mathrm{A}}{\partial \mathbf{x}_\mathrm{A}} 
        - \mathbf{G}_\mathrm{A} \mathbf{x}_\mathrm{A} - \mathbf{C} \mathbf{u}_\mathrm{S},\\
        \label{eq:eom_2B}
        \ddot{\mathbf{u}}_\mathrm{S} = - \mathbf{C}^T \mathbf{x}_\mathrm{A} - \pmb{\omega}^2 \mathbf{u}_\mathrm{S} ,
    \end{gather}
    where $\mathbf{C}=\mathbf{G}_\mathrm{AS}\mathbf{U}$ is a matrix of couplings between each adsorbate coordinate and solid phonon mode, and $\pmb{\omega}^2=\mathbf{U}^T \mathbf{H}_\mathrm{S} \mathbf{U}$ is a diagonal matrix of the squared frequencies of the solid phonons.
    Combining Eq.~\ref{eq:eom_2A} and Eq.~\ref{eq:eom_2B} yields a GLE for the adsorbate degrees of freedom\cite{zwanzig_nonequilibrium_2001, tuckerman_statistical_2023},
    \begin{equation}
        \label{eq:gle_1}
        \ddt{\mathbf{x}}_\mathrm{A} =  
        -\frac{\partial V_\mathrm{A}}{\partial \mathbf{x}_\mathrm{A}} 
        - \left[ \mathbf{G}_\mathrm{A} - \mathbf{K}(t=0) \right] \mathbf{x}_\mathrm{A}(t)
        - \int_0^t \mathbf{K}(t-\tau) \dt{\mathbf{x}}_\mathrm{A}(\tau) d\tau + \mathbf{R}(t),
    \end{equation}
    where the friction kernel $\mathbf{K}(t)$ is given by,
    \begin{equation}
        \label{eq:mem_1}
        \mathbf{K}(t) = \mathbf{C} \frac{ \cos(\pmb{\omega} t) }{\pmb{\omega}^2} \mathbf{C}^T,
    \end{equation}
    and the statistics of the random force $\mathbf{R}(t)$ are related to $\mathbf{K}(t)$ by the second fluctuation-dissipation theorem,
    \begin{equation}
        \label{eq:randomF_1}
        \frac{\left \langle  \mathbf{R}(t) \mathbf{R}^T(0)\right \rangle}{k_{\mathrm{B}} T} = \mathbf{K}(t).
    \end{equation}
    The positive frequency components of the Fourier transform of the friction kernel are (up to a multiplicative constant),
    \begin{equation}
        \label{eq:spectral_1}
        \bar{K}_{ab}(\omega) = \sum_{j} \frac{C_{aj}C_{bj}}{\omega^2_j} \delta{(\omega - \omega_j)},
    \end{equation}
    where $\omega_j$ is the frequency of $j$th phonon mode of the solid.
    Alternatively, we can express this quantity as, 
    \begin{equation}
        \label{eq:spectral_2}
        \bar{K}_{ab}(\omega) = \frac{C_a(\omega) C_b(\omega)}{\omega^2} \rho(\omega),
    \end{equation}
    where $\rho(\omega)$ is the density of phonon modes at frequency $\omega$.

    Some insight can be derived by studying the formulation above.
    Equation~\ref{eq:gle_1} illustrates that within the harmonic approximation, the influence of phonon modes on adsorbates can be described entirely in terms of $\mathbf{K}(t)$, as it determines both the properties of $\mathbf{R}(t)$ via the fluctuation-dissipation theorem and the deviation of the adsorbate potential from that of an adsorbate on a fixed solid. 
    Equations~\ref{eq:spectral_1} and~\ref{eq:spectral_2} reveal that the Fourier transform of $\mathbf{K}(t)$ can be interpreted as the phonon density of states reweighted by the value of the adsorbate-phonon coupling $\mathbf{C}$. 
    Indeed, $\mathbf{\bar{K}}_{ab}(\omega)$ is often referred to as the \textit{spectral density} of the environment. 
    
    In some cases Eqs.~\ref{eq:mem_1} and~\ref{eq:spectral_1} may further be simplified by stating that adsorbates couple only to a handful of surface sites and thus the matrices $\mathbf{G}_{AB}$ and $\mathbf{C}$ are sparse. For example, consider the case of an adsorbate that interacts with a single surface atom via a single adsorbate atom (e.g. CO on Pt(111) \cite{ertl_chemisorption_1977,steininger_adsorption_1982}). If the force constant associated with this interaction is $\mu \omega_{\mathrm{as}}^2$, where $\mu$ is the reduced mass of the adsorbate and surface atom, Eqs.~\ref{eq:mem_1} and \ref{eq:spectral_1} may be simplified to, 
    \begin{equation}
        \label{eq:mem_2}
        K(t) =  \frac{\mu^2}{m M} \omega_{\mathrm{as}}^4 \sum_{j} \frac{ U^2_{sj}}{\omega^2_j} \cos(\omega_j t),
    \end{equation}
    \begin{equation}
        \label{eq:spectral_3}
        \bar{K}(\omega) = \frac{\mu^2}{m M}  \omega_{\mathrm{as}}^4 \sum_{j} \frac{ U^2_{sj}}{\omega^2_j} \delta{(\omega - \omega_j)},
    \end{equation}
    where $U_{sj}$ is the expansion coefficient of the surface displacement $s$ in the $j$th normal mode, $m$ is the mass of the adsorbate atom, and $M$ is the mass of the surface atom. 

    The formulation above provides a framework for representing the influence of surface phonons on an adsorbate implicitly via the GLE.
    With a well-defined model of an adsorbate, surface, and their interactions, the friction kernel can be calculated. 
    In the following subsection, we illustrate the application of this formalism to simple model systems.
    
    \subsection{Results and discussion}
    \label{subsec:sec2_results}

    We have computed the phonon friction kernel for a simple model of CO on Pt(111). 
    The CO was modeled as a single adatom and assumed the interact with a single adsorption site, as experimental structures show that CO adsorbs primarily atop Pt(111) sites \cite{ertl_chemisorption_1977,steininger_adsorption_1982}. 
    We artificially vary the adsorbate-surface frequency $\omega_{\mathrm{as}}$ over a range of physically motivated values, in order asses how the strength of the adsorbate-surface interaction affects the properties of the friction kernel.
    The metal potential energy surface $V_S$ was modeled using an effective medium theory (EMT) forcefield developed by Norskov et al.\cite{Norskov1980}
    More details about calculations can be found in the Section~\ref{sec:methods}. 
    
    \begin{figure*}[h]  
       \centering
       \includegraphics*[width=5.25in]{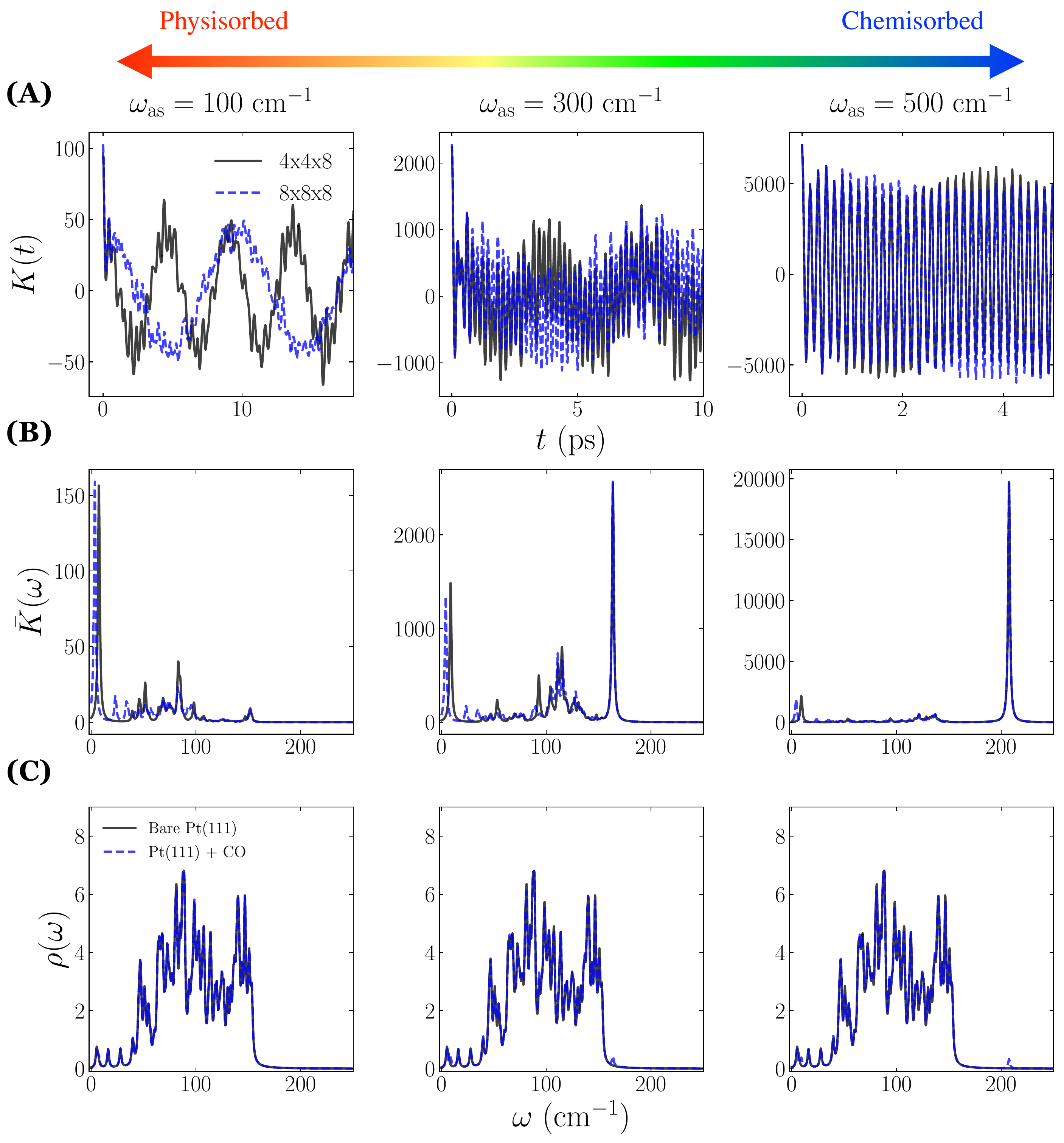}
       \caption{ (A) Friction kernel, (B) spectral density, and (C) density of states for Pt(111) and three values of $\omega_{\mathrm{as}}$. The density of states in (C) was calculated using the 4x4x8 surface slab.}
       \label{figure1}
    \end{figure*}
    
    Fig.~\ref{figure1} illustrates the results of our calculations for a selected set of $\omega_\mathrm{as}$ values ranging from those characteristic of weakly adsorbed species to those characteristic of strongly adsorbed species.  
    We plot the friction kernel computed in the direction normal to the surface in systems with either a 4x4x8 or 8x8x8 Pt(111) slab.
    The lower values of $\omega_\mathrm{as}$ that we consider lie below that of the platinum's Debye frequency of $\omega_{\mathrm{D}} = 156 \mathrm{cm}^{-1}$.
    We associate these values with the physisorbed regime, as they are characteristic of weakly adsorbed systems such as noble gases on Pt(111).\cite{chen_role_2012}
    The higher values of $\omega_\mathrm{as}$ that we consider are associated with the chemisorbed regime, corresponding to physical systems such as CO on Pt(111), which has $\omega_\mathrm{as} \approx 480\ \mathrm{cm}^{-1}$).\cite{steininger_adsorption_1982,lahee_low_1986,gunasooriya_co_2018}
    Delta function peaks in $\bar{K}(\omega)$ and $\rho(\omega)$ were broadened to thin Lorentzians of width $1 \mathrm{cm}^{-1}$ for ease of visualization.
    \textcolor{black}{It is important to note that the non-decaying oscillations in $K(t)$ are a consequence of the harmonic approximation. Accounting for anharmonicities will cause the memory kernel to asymptotically decay at long-times.}
    
    For all cases of $\omega_{\mathrm{as}}$, the adsorbate couples strongly to a surface acoustic mode appearing around $\omega = 10\mathrm{cm}^{-1}$. 
    The location of this peak is highly dependent on the dimensions of the simulated surface slab, consistent with the behavior of an acoustic phonon. 
    In Sec.~\ref{sec:dispersion}, we give a more detailed theoretical and numerical analysis of the size effects observed here, fully accounting for phonon dispersion. 
    In the physisorbed regime, the friction kernel is dominated by this acoustic mode, resulting in highly non-Markovian behavior.
    The phonon mode associated with this peak is associated with the flexing of the lattice in the direction perpendicular to the surface. 
    Snapshots of this motion are provided in the insets in Fig.~\ref{figure_intro} and video animations are provided in the electronic SI. 

    In the chemisorbed regime, the friction kernel is dominated by a high frequency mode around $\omega = 210 \ \mathrm{cm}^{-1}$. This mode arises from the local oscillations of the surface site the adsorbate is bound to. 
    The frequency of this mode is independent of simulation cell size, suggesting that it is dispersionless.
    The large amplitude of this mode signifies that in the chemisorbed limit, the adsorbate is primarily sensitive to the local oscillations of the surface binding site (which are shifted in frequency due to the presence of the adsorbate). 
    Many published models for reactive scattering on solid surfaces will describe solid vibrations using only a single, harmonically bound surface atom \cite{luntz_ch4_1991,nattino_modeling_2016,rittmeyer_energy_2018,zhou_modified_2019,zhou_dynamics_2020}. These results explain why such a method is successful for strongly coupled species. 
    
    Despite this large shift in the spectral density when varying $\omega_{\mathrm{as}}$, in Fig~\ref{figure1}C we demonstrate that the phonon density of states is nearly identical to that of a bare surface. 
    The only significant change in $\rho(\omega)$ is the presence of the aforementioned surface site local mode present around $\omega = 210 \ \mathrm{cm}^{-1}$. 
    These results make physical sense given that the low frequency acoustic modes of a solid should be unaffected by gaseous species, especially at low pressures and surface coverages. 
    
    To evaluate the effect of the solid force field and structure on these results, we have tested several different crystal structures, facets, elemental compositions, and force fields. 
    The results of these tests are presented in the Section S1 of the SI. 
    While the quantitative properties of the friction kernel vary across different systems (e.g. Ru has a much higher Debye frequency than Pt), the qualitative dependence of the friction kernel on $\omega_\mathrm{as}$ is quite general.
    This independence with respect to the details of the atomistic model is appealing and can be understood from a perturbative perspective. 
    Specifically, in the physisorbed limit, the motion of the surface binding site is a small perturbation to the bulk phonon modes, while in the chemisorbed limit, the bulk phonon modes are a small perturbation to the motion of the surface binding sites. 
    In the SI, we rigorously examine this statement by comparing perturbative schemes to the exact results for $K(t)$ and $\bar{K}(\omega)$.
    \textcolor{black}{Perturbation theory also allows us to derive analytical approximations to the friction kernel (Eqs. S10-S12) and precisely define the "phase boundaries" in Figure~\ref{figure_intro}.}

    We acknowledge that in this section we have only studied the effects of phonons for an ideal clean surface.
    Presumably, sources of surface heterogeneity, such as steps and defects, may add further richness and complexity to the picture we have provided here. 
    We also note that we have only studied elemental solids, and the optical modes of polyatomic crystals could add another interesting dimension to our physical picture. 
    \textcolor{black}{Finally, in this section we have only studied species at a gas-solid interface. Many important surface-chemical processes occur at liquid-solid interfaces, where it has been demonstrated sorption dynamics play a critical role in electrokinetic transport and measurements~\cite{mangaud_chemisorbed_2022}}.

\section{Phonon dispersion}
    \label{sec:dispersion}
    The dependence of the acoustic peak frequency on simulation size, such as illustrated in Fig.~\ref{figure1}, is an artifact that has the potential to prevent straightforward comparison between experiment and simulation. 
    With or without periodic boundary conditions, the size of the simulated solid limits the maximum wavelength surface phonon.
    In this section, we illustrate how to generalize the theory presented in Section~\ref{sec:phonon_induced_friction} by computing the friction kernel in the limit of an infinite surface through integration of frequencies and eigenmodes across the first Brillouin zone.  

\begin{figure*}[h]  
       \centering
       \includegraphics*[width=5.25in]{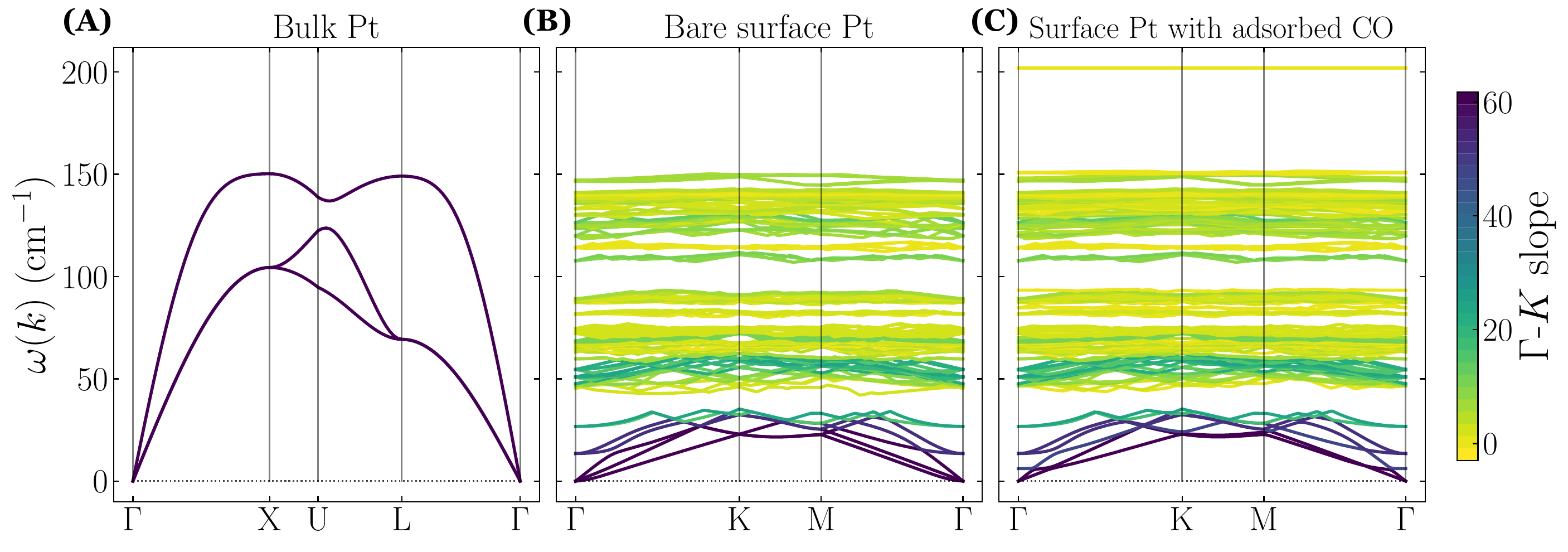}
       \caption{Phonon dispersion curves. (A) Bulk Pt dispersion curves calculated using an EMT forcefield and a 10x10x10 atom supercell. (B) Dispersion curves for a 4x4x8 atom surface slab replicated in a 6x6 surface cell (C) Same as middle but with an adsorbed CO molecule corresponding to $\omega_{\mathrm{AS}}=480\ \mathrm{cm}^{-1}$.}
       \label{figure3}
    \end{figure*}

    In an infinite crystalline solid, displacement by a lattice vector returns the same solid. 
    This symmetry can be leveraged to calculate the phonon frequencies and displacements of the bulk solid using a spatial Fourier transform of the mass-weighted Hessian. 
    This approach leads to the well-known Bloch's theorem, which is summarized as follows. 
    Let $a$ and $b$ be the indices of two primitive unit cells, and let $\mathbf{H}_S(a,b)$ be the mass-weighted Hessian of the crystal, ${H}_{S;ij}(a,b)=\frac{\partial^2 V}{\partial {x}_{S,i}(a)  \partial {x}_{S,j}(b)}$. The Fourier transform of this matrix may be expressed as,
    \begin{equation}
        \mathbf{D}(\mathbf{k}) = \sum_{a,b} \mathbf{H}_\mathbf{S}(a,b) e^{i\mathbf{k} \cdot (\mathbf{r}_a-\mathbf{r}_b)},
    \end{equation}
    where $\mathbf{r}_a$ is the origin of the $a$th cell. $\mathbf{D}(\mathbf{k})$ is known as the \textit{dynamical matrix} satisfying,
    \begin{equation}
        \mathbf{D}(\mathbf{k}) \mathbf{U}_j(\mathbf{k}) = -\omega_j^2(\mathbf{k}) \mathbf{U}_j(\mathbf{k}).
    \end{equation}
    where $\omega_{j}$ is the $j$th phonon band frequency and $\mathbf{U}_j$ is the corresponding polarization vector. 
    For 3D monatomic crystals, the primitive (Wigner-Seitz) cell consists of a single atom with three degrees of freedom, thus producing three phonon bands. 
    These bands, illustrated in Fig.~\ref{figure3} for FCC platinum, are the acoustic transverse and longitudinal modes, and all characterized by linear dispersion at low wavenumber. 
    
    In a surface slab, the presence of anisotropy breaks the symmetry of the 3D crystal. 
    This symmetry breaking leads to additional surface mode bands, which can be acoustic (Rayleigh waves), or non-acoustic in character\cite{hong_first-principles_2005,benedek_theory_2010,bortolani_surface_1989}. In Fig. 4B. and 4C. we demonstrate the phonon dispersion of a Pt(111) surface calculated via EMT. These results were calculated using a 4x4x8 surface replicated in a 6x6 super-cell. We verified convergence with respect to supercell size. 
    Of course, the number of surface phonon bands depends on the size of the surface unit cell used in computing $\mathbf{D}(\mathbf{k})$, however, we demonstrate in in Fig.~S7 results are qualitatively similar across different unit cell sizes. 

    Figure~\ref{figure3}B illustrates the phonon dispersion for a bare surface, while Figure~\ref{figure3}C illustrates the dispersion for a surface with an adsorbed CO. The bands are colored based on their mean slope between the $\Gamma$ and $K$ high-symmetry points. 
    The three lowest frequency bands are all surface acoustic modes, as exemplified by the flexing mode depicted in the inset of Fig.~\ref{figure_intro}, confirming that the corresponding peak in the phonon spectral density in Fig.~\ref{figure1} arises from an acoustic phonon. 
    The remaining modes are nearly dispersionless --- especially the highest frequency mode in Fig.~\ref{figure3}C, which corresponds to the surface site local vibration. The dispersionless nature of this mode supports why it was not seen to be dependent on surface slab size in Fig.~\ref{figure1}. 
    
    Using these dispersion relations, we can average the friction kernel across the first Brillouin zone,
    \begin{equation}
        \label{eq:mem_kspace}
        K(t) =   \frac{\mu^2}{mM} \omega_{\mathrm{as}}^4 \sum_j {\sum_{\mathbf{k}}}' \
        \frac{ \left| U_{sj} (\mathbf{k}) \right|^2 }{\omega_j(\mathbf{k})^2 } \cos( \mathbf{\omega}_j(\mathbf{k}) t)
         ,
    \end{equation}
    \begin{equation}
        \label{eq:spectral_kspace}
        \bar{K}(\omega) =  \frac{\mu^2}{mM} \omega_{\mathrm{as}}^4 \sum_j {\sum_{\mathbf{k}}}' \ 
        \frac{ \left| U_{sj} (\mathbf{k}) \right|^2 }{\omega_j(\mathbf{k})^2 } \delta( \omega - \mathbf{\omega}_j(\mathbf{k}))
         ,
    \end{equation}
    where the primed summation is taken over all wavevectors in the first Brillouin zone and the sum over $j$ is taken over all surface unit cell phonon modes. 
    Equations~\ref{eq:mem_kspace} and \ref{eq:spectral_kspace} simply separate the sum in Eqs.~\ref{eq:mem_2} and \ref{eq:spectral_3} into two parts: the outer sum varying intra-cell displacements and the inner sum varying inter-cell displacements. 
    In principle, this procedure eliminates the dependence on boundary conditions and generalizes results from a surface unit cell to an infinite periodic surface. 
    However, the depth of the surface (corresponding to the non-periodic dimension) is still limited to the depth of the surface unit cell. 

    \begin{figure*}[h]  
       \centering
       \includegraphics*[width=5.25in]{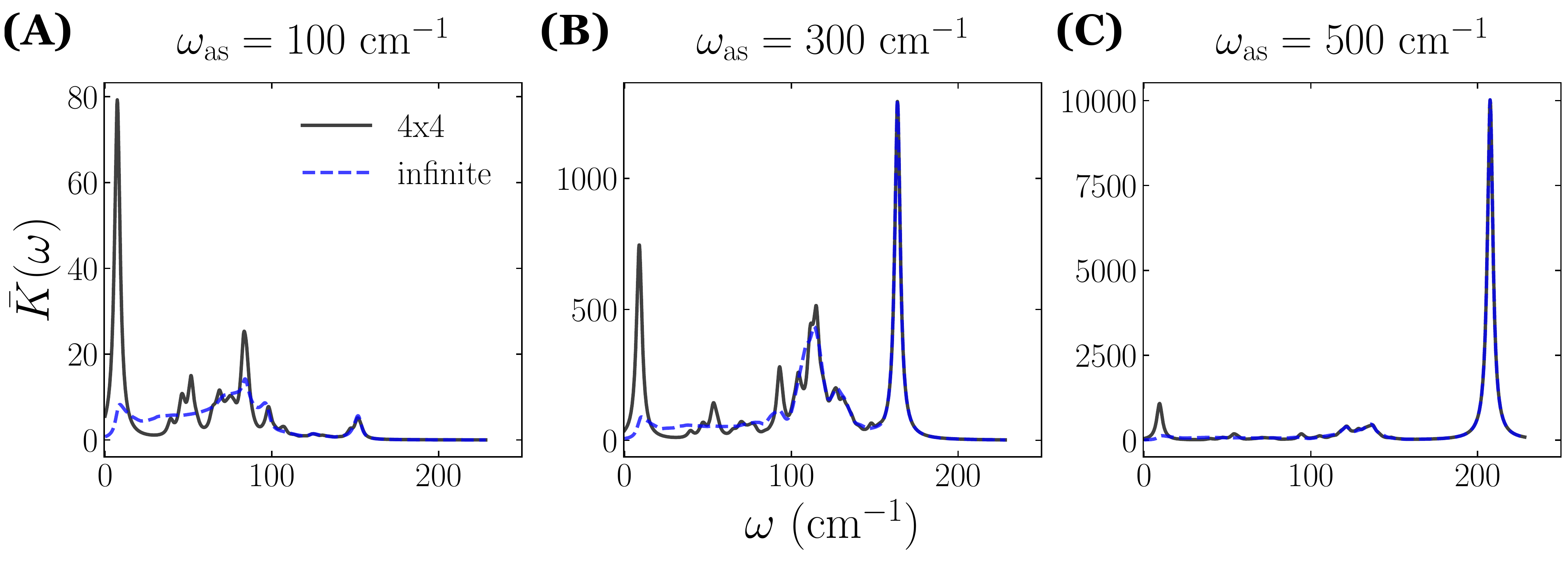}
       \caption{ Friction kernel for a model adsorbate on Pt(111) in a finite system (black) and in the infinite system size limit (blue).
       The finite system is equivalent to the 4x4x8 results presented in Fig.~\ref{figure1}. 
       }
       \label{figure4}
    \end{figure*}
    Figure~\ref{figure4} demonstrates the results of Eqs.~\ref{eq:mem_kspace} and \ref{eq:spectral_kspace} and compares to previous results for a single unit cell slab. 
    Naturally, the strong coupling of the adsorbate to a single acoustic mode is broadened, resulting in a much flatter spectral density in the low frequency ranges.
    Such a flat spectral density is characteristic of Markovian (white) noise. 
    The high frequency range of the spectral density is largely unaltered, due to the dispersionless nature of the high frequency modes.
    In the SI, we discuss how a flat spectral density for acoustic phonons is consistent with predictions from continuum elastic theory. 
    
\section{Phonon effects on reaction rates}
\label{sec:rates}
    One of the primary motivations for studying the atomic vibrations of a solid surface is to elucidate the role of phonons in reactions rates at catalytic interfaces. 
    As was noted first by Kramers\cite{kramers_brownian_1940}, and expanded on by many others\cite{grote_stable_1980,carmeli_nonmarkovian_1983,pollak_theory_1989,kappler_memory-induced_2018}, the friction of the environment plays a critical role in determining reaction rates, and especially in determining rate prefactors. 
    In this section, we use transition state theory (TST) to demonstrate how surface phonons affect reaction rates, emphasizing the role of the phonon friction kernel. 
    We then specialize to the case of desorption rates and compare the results of our model to experimental temperature dependent rate constants for CO and Xe desorption from Pt(111).
    Several details of the formalism given this section have been omitted for the purpose of providing a concise presentation of the theory that focuses on physical ramifications. 
    A thorough presentation of the theory is provided in the SI for interested readers. 

    \subsection{Transition State Theory}
    \subsubsection{General theory}
    The Eyring transition state theory (TST) equation for the rate constant may be expressed as,
    \begin{equation}
        \label{eq:k_Eyring}
        k = \frac{1}{\beta h} \frac{Q^{\ddag}}{Q_R} e^{-\beta \Delta E^{\ddag}},
    \end{equation}
    where $h$ is Planck's constant, $\Delta E^{\ddag}$ is the potential energy difference between the reactant state and the transition state, $Q^{\ddag}$ is the partition function at the transition state, and $Q_R$ is the partition function in the reactant basin. 
    By expanding said partition functions to second order about the transition state and the reactant basin, respectively, we can generate analytical expressions for $Q_R$ and $Q^{\ddag}$ yielding,
    \begin{equation}
        \label{eq:k_harmonic_TST}
        k = \frac{1}{2 \pi}
        \times
        \frac{\prod\limits^{N-1}_{i=0} f_i}{\prod\limits^{N-1}_{i=1} f_i^{\ddag}}
        \times
        e^{-\beta \Delta E^{\ddag}},
    \end{equation}
    where $N$ is the total number of modes, $f_i$ are the vibrational frequencies in the reactant basin, $f_i^{\ddag}$ are the \textit{non-imaginary} vibrational frequencies at the transition state. 
    Note that since the transition state is defined to be a saddle point, it always has one imaginary frequency mode and this mode corresponds the reaction coordinate. 
    In a gas phase molecular reaction, $f_i$ is the $i$th eigenfrequency of the molecular mass-weighted Hessian. However, in a reaction at a surface interface, both surface phonon modes and molecular/adsorbate modes will contribute to the rate constant.
    We thus separate the product of eigenfrequencies into contributions from the solid surface and the  molecular adsorbate to yield,
    \begin{equation}
        \label{eq:k_harmonic_phonon}
        k = \frac{1}{2 \pi} 
        \times
        \prod\limits^{N_S-1}_{i=0} \frac{\omega_i}{\omega_i^{\ddag}} 
        \times
        \frac{\prod\limits^{N_A-1}_{i=0} \tilde{f}_i}{\prod\limits^{N_A-1}_{i=1} \tilde{f}^{\ddag}_i}
        \times
        e^{-\beta \Delta E^{\ddag}},
    \end{equation}
    where $N_S$ and $N_A$ are the number of solid and adsorbate degrees of freedom respectively ($N_S + N_A = N)$), $\omega_i$ and $\omega_i^{\ddag}$ are the phonon frequencies in the reactant and transition state respectively, and $\tilde{f}_i$ and $\tilde{f}^{\ddag}_i$ are effective molecular frequencies in the reactant and transition state respectively. 
    Let $\mathbf{H}_A$ be the mass-weighted Hessian of the adsorbate degrees of freedom, then $\tilde{f}_i$  and $\tilde{f}^{\ddag}_i$ are the eigenfrequencies of an effective Hessian,
    \begin{equation}
        \label{eq:eff_hess_A}
        \tilde{\mathbf{H}}_A = \mathbf{H}_A - \mathbf{C}\pmb{\omega}^{-2}\mathbf{C}^T,
    \end{equation}
    where the shift term is equal to the instantaneous ($t=0$) friction, $\mathbf{K}(t=0)=\mathbf{C}\pmb{\omega}^{-2}\mathbf{C}^T$.

    Comparing Eqs. \ref{eq:k_harmonic_TST} and \ref{eq:k_harmonic_phonon} reveals that phonons introduce two distinct corrections to rate constant: the ratio of solid phonon frequencies between the reactant and transition state, and the shift in adsorbate frequencies. 
    The ratio of phonon frequencies is largest when the reactant molecules are strongly coupled to the surface and the transition state is not. In Fig.~\ref{figure1}C we demonstrated that an adsorbate strongly coupled to a surface affects only the highest frequency phonon mode, leaving the bulk of the phonon density of states unchanged. Thus, we can approximate the ratio of phonon frequencies as,
    \begin{equation}
        \label{eq:phonon_ratio_approx}
        \prod\limits^{N_S-1}_{i=0} \frac{\omega_i}{\omega_i^{\ddag}} 
        \approx
        \frac{\tilde{\omega}_{\mathrm{D}}}{\omega_{\mathrm{D}}},
    \end{equation}
    where $\tilde{\omega}_{\mathrm{D}}$ is the highest frequency phonon mode when the reactants are bound to the surface and $\omega_{\mathrm{D}}$ is the bare solid Debye frequency. \textcolor{black}{The factor of $\tilde{\omega}_{\mathrm{D}}$ is directly related to our observation that chemisorbed species couple strongly to local modes above the natural Debye frequency of the metal. The outsized influence of these modes on the dynamics of the chemisorbed species reflects why it appears in the expression for the rate constant. For physisorbed species $\frac{\tilde{\omega}_{\mathrm{D}}}{\omega_{\mathrm{D}}} = 1$.}
    
    The shift in the adsorbate frequencies is a thermodynamic correction arising from phonons altering the free energy surface along the reaction coordinate. Indeed, if we denote $\tilde{f}_0$ as the adsorbate normal mode along the reaction coordinate Eq.~\ref{eq:k_harmonic_phonon} can be simplified to,
    \begin{equation}
        \label{eq:k_harmonic_phonon2}
        k = \frac{\tilde{f}_0}{2 \pi} 
        \times
        \prod\limits^{N_S-1}_{i=0} 
        \frac{\omega_i}{\omega_i^{\ddag}} 
        \times
        e^{-\beta (\Delta E^{\ddag} + T \Delta \tilde{S}^\ddag)},
    \end{equation}
    where $\Delta \tilde{S}^\ddag$ is the effective barrier entropy,
    \begin{equation}
        \label{eq:S_barrier}
        \Delta \tilde{S}^\ddag = k_{\mathrm{B}} 
        \sum^{N_A-1}_{i=10} \ln \left( \frac{\tilde{f}^{\ddag}_i}{\tilde{f}_i}  \right).
    \end{equation}
    In the harmonic approximation this barrier entropy is independent of temperature, but in a more general context it can be shown to be temperature dependent. 

    \textcolor{black}{
    In Eqs. \ref{eq:k_harmonic_phonon} and \ref{eq:k_harmonic_phonon2}, the influence of the phonon friction kernel on the rate constant is both explicit and implicit. It is explicit in terms of $K(t=0)$, but also implicit in ratio of normal mode frequencies. The dependence of this ratio on the friction kernel can only be made explicit if additional simplifications are made. Notably, under the condition that the coupling between the adsorbate degrees of freedom and the solid degrees of freedom $\mathbf{C}$ is the same in both the reactant basin and transition state, harmonic TST reduces to Kramers-Grote-Hynes(KGH) theory, where the rate prefactor is explicitly related to the Laplace transform of the friction kernel \cite{pollak_theory_1986}. However, such a condition is not appropriate even in simple desorption processes, and therefore we do not directly use KGH theory here. See the SI (Section S6) for additional details on KGH theory. 
    }
    
    \subsubsection{Surface desorption}
    The formalism described in the previous subsection can be applied to derive a rate constant for surface desorption.
    For gas-phase desorption, the reaction coordinate can be defined as the distance of the adsorbate center of mass from the surface.
    Furthermore, in many cases, the desorption process is \textit{barrierless}, meaning that at the transition state the molecule and surface do not interact \cite{tully_dynamics_1994}.
    Thus, we can adapt Eq.~\ref{eq:phonon_ratio_approx} to express the gas-phase desorption rate constant,
    \begin{equation}
        \label{eq:k_desorption1}
        k_d = \frac{\tilde{\omega}_{\mathrm{as}}}{2 \pi}
        \times
        \frac{\tilde{\omega}_{\mathrm{D}}}{\omega_{\mathrm{D}}}
        \times
        e^{-\beta \Delta E^{\ddag} },
    \end{equation}
    where $\tilde{\omega}_{\mathrm{as}} $ is the effective adsorbate-surface interaction frequency satisfying,
    \begin{equation}
        \tilde{\omega}_{\mathrm{as}} = \sqrt{\frac{\mu}{m} \omega^2_{\mathrm{as}} - K(t=0)}.
    \end{equation}
    
    While transition state theory is primarily a classical theory \cite{garrett_generalized_1979,pollak_reaction_2005,chandler_statistical_2008,miller_journey_2014}, studies have demonstrated improved agreement with experiment when introducing simple quantum corrections, such as accounting for the rotational motion of the molecule or using quantum harmonic oscillator partition functions instead of classical oscillator partition functions \cite{tully_dynamics_1994,golibrzuch_co_2015}. 
    We will thus compare four different models for the desorption rate constant to experimental results:
    (1) a fixed-surface model using classical harmonic oscillator partition functions and a rotation correction,
    \begin{equation}
        \label{eq:kd1}
        k_{d1} = 
        \frac{\omega_{\mathrm{as}}}{2 \pi} 
        \times
        \frac{2 I }{\hslash^2 \beta}
        \times
        e^{-\beta \Delta E^{\ddag} },
    \end{equation}
    where $I$ denotes the moment of inertia of the adsorbate. 
    (2) a phonon-corrected model using classical harmonic oscillator partition functions and a rotational correction,
    \begin{equation}
        \label{eq:kd2}
        k_{d2} = 
        \frac{\tilde{\omega}_{\mathrm{as}}}{2 \pi} 
        \times
        \frac{\tilde{\omega}_{\mathrm{D}}}{\omega_{\mathrm{D}}}
        \times
        \frac{2 I }{\hslash^2 \beta}
        \times
        e^{-\beta \Delta E^{\ddag} },
    \end{equation}
    (3) a fixed-surface model using quantum harmonic oscillator partition functions and a rotational correction,
    \begin{equation}
        \label{eq:kd3}
        k_{d3} = 
        \frac{1-e^{-\beta \hslash \omega_{\mathrm{as}}}}{2 \pi \beta \hslash}
        \times
        \frac{2 I }{\hslash^2 \beta}
        \times
        e^{-\beta \Delta E^{\ddag} },
    \end{equation}
    and (4) a phonon-corrected model using quantum harmonic oscillator partition functions and a rotational correction,
    \begin{equation}
        \label{eq:kd4}
        k_{d4} = 
        \frac{1-e^{-\beta \hslash \tilde{\omega}_{\mathrm{as}}}}{2 \pi \beta \hslash}
        \times
        \frac{1-e^{-\beta \hslash \tilde{\omega}_{\mathrm{D}}}}{1-e^{-\beta \hslash \omega_{D}}}
        \times
        \frac{2 I }{\hslash^2 \beta}
        \times
        e^{-\beta \Delta E^{\ddag} }.
    \end{equation}
   Note that by "fixed-surface" we do not mean a surface at absolute zero, but rather a surface that acts as an ideal, structureless, thermal environment. 
    
    \subsection{Results and discussion}
    We have compared results from Eqs.~\ref{eq:kd1} to ~\ref{eq:kd4} to experimental temperature-dependent desorption rate constants for CO and Xe from a Pt(111) surface. 
    Parameters used in computing Eqs.~\ref{eq:kd1} to ~\ref{eq:kd4} are presented in Table~\ref{tab:rate_constant}. We ignore the rotational partition function factors for Xe calculations. 
    The phonon corrections in Eq.~\ref{eq:kd2} and Eq.~\ref{eq:kd4} were computed using a 4x4x8 EMT surface slab and subsequently averaged across the first Brillouin zone to approximate an infinite surface, as described in Sec.~\ref{sec:dispersion}. 
    In the SI, we illustrate that the rate constant corrections we present here are not sensitive to the size of the surface slab used, or whether one accounts for phonon dispersion. 

    \begin{table}[htb]
        \caption{Parameters used for computing desorption rate constants shown in Fig.~\ref{figure5}.}
        \label{tab:rate_constant}
        \begin{tabularx}{3.25in}{XXXXX}
            \toprule
            & $E^{\ddag}$ ($\mathrm{eV}$) & $\omega_{\mathrm{as}}$ ($\mathrm{cm^{-1}}$) & $\tilde{\omega}_{\mathrm{as}}$ ($\mathrm{cm^{-1}}$) & $\tilde{\omega}_{\mathrm{D}}$ ($\mathrm{cm^{-1}}$) \\ 
            \midrule
            CO & 1.47\cite{golibrzuch_co_2015}  & 480\cite{steininger_adsorption_1982} & 164 & 203 \\
            Xe & 0.245\cite{rettner_measurement_1990} & 28\cite{chen_role_2012} & 21 & 156 \\
            \bottomrule
        \end{tabularx}
    \end{table} 
    
    The CO desorption rate constants were taken from Ref.~\citenum{golibrzuch_co_2015} and the Xe desorption rates were taken from Ref.~\citenum{rettner_measurement_1990}. In Ref.~\citenum{golibrzuch_co_2015}, desorption rate constants were calculated by fitting the time-dependent flux from a beam scattering experiment to two models: a single exponential model and a bi-exponential model. The single exponential model fit the entire flux signal, mixing contributions from terrace and steps. Meanwhile the biexponential model separated the flux into a fast component, arising from terrace desorption, and a slow component, arising from step to terrace diffusion followed by terrace desorption. While our TST calculations do not include the role of steps, we compare the results of our models to data both from the single exponential model and the fast component of the biexponential for thoroughness and transparency. 
    
    \begin{figure}[h]  
       \centering
       \includegraphics*[width=2.75in]{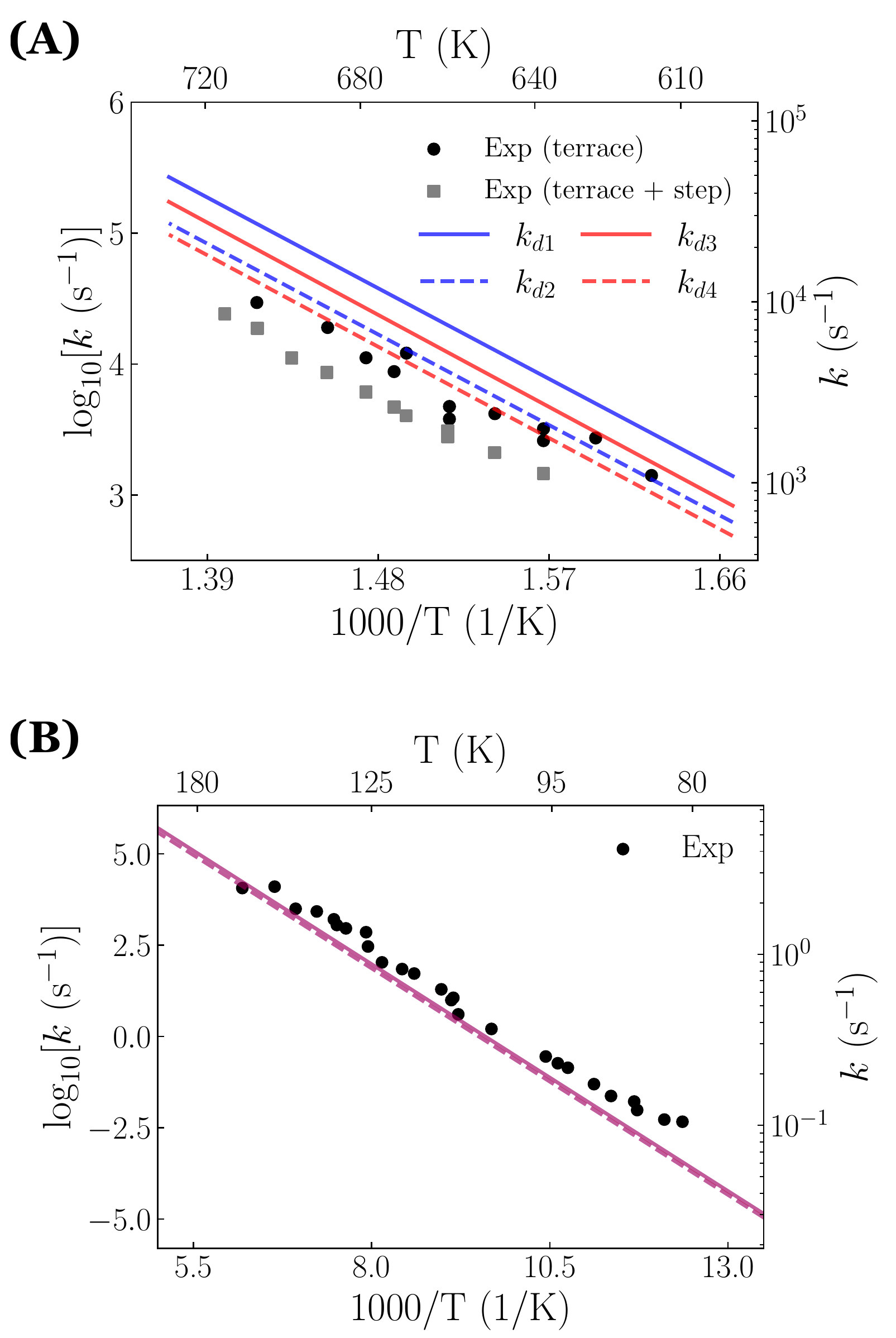}
       \caption{ Rate constants for desorption from a Pt(111) surface. (A) CO desorption. Grey squares are experimental data which mixed contributions from both steps and terraces. Black circles refer experimental data where the kinetics of terrace desorption was isolated. (B) Xe desorption.}
       \label{figure5}
    \end{figure}
    In Fig.~\ref{figure5}A, we see that the phonon-corrected models (Eqs.~\ref{eq:kd2} and ~\ref{eq:kd4}) give improved agreement with experimental results for CO desorption. In particular, $k_{d4}$ the quantum, flexible surface model and the terrace desorption rate constants yield the best agreement. The improved agreement when using Eqs.~\ref{eq:kd2} to ~\ref{eq:kd4} versus Eqs.~\ref{eq:kd1} to ~\ref{eq:kd3} arises from the reduced adsorbate-surface frequency $\tilde{\omega}_\mathrm{as}$. 
    Physically, the flexible surface reduces the stiffness of the adsorbate-surface bond, leading to a lower frequency of attempts over the barrier and a lower rate prefactor. 
    
    In Fig.~\ref{figure5}B, we demonstrate results for Xe. Here, all the TST models lie essentially on top of each other and are lower in value than the experimental rate constants, although by a small margin. The smaller phonon corrections for Xe versus CO are a natural result of the weaker interactions with the surface. Weaker coupling means a small $K(t=0)$, leading to $\tilde{\omega}_{\mathrm{as}} \approx \omega_{\mathrm{as}}$. Furthermore, weak coupling also results in a phonon density of states that is unchanged from a bare lattice, implying $\tilde{\omega}_{\mathrm{D}} = \omega_{\mathrm{D}}$. In general, the theory presented in this section suggests the phonon corrections to the rate constant are much larger for chemisorbed species than physisorbed species. \textcolor{black}{We also note that the deviation between theory and experiment in Fig.~\ref{figure5}B is larger for lower temperatures. For sufficiently low-mass species at sufficiently cryogenic temperatures, it has been shown that tunneling plays a significant role in barrier crossing\cite{hanggi_reaction-rate_1990}, requiring more sophisticated quantum corrections to the rate constants than those in Eq.~\ref{eq:kd3} and Eq.~\ref{eq:kd4}.}

    It is worth emphasizing that using a slightly different values for the surface binding energy, $\Delta E^{\ddag}$, can substantially shift the quality of agreement of theoretical calculations with experiment. The major impediment to the first principles calculation of chemical rates is still the calculation of the barrier energy, and the phonon corrections to the rate constants seem to be a comparatively minor factor, even for chemisorbed species. Indeed, the purpose of this section was not to demonstrate that the magnitude of phonon corrections to reaction rates is large, but rather to illustrate that the theoretical models we developed in Sections 2 and 3, when combined with transition state theory, produce physically interpretable results which correspond well with existing experimental measurements. 

\section{Methods}
\label{sec:methods}
    All calculations were performed using the Atomic Simulation Environment (ASE) python package \cite{larsen_atomic_2017,tadmor:elliott:2011,OpenKIM-MO:108408461881:001,OpenKIM-MD:128315414717:004,OpenKIM-MO:757342646688:000,OpenKIM-MD:120291908751:005}. Mass-weighted Hessians were computed using a finite-difference scheme in the vibrations module of ASE. Dynamical matrices were computed using the phonon module. The displacement size used in finite difference calculations was 0.01. 100 points were used to uniformly sample the first Brillouin zone. 

    When computing the friction kernels and spectral densities, all adsorbates were treated as an effective adatom, with a given coupling between the center of mass of the adsorbate and a surface atom. Friction kernels were computed for the three Cartesian degrees of freedom of this adatom. Additionally, a single atom in the bottom layer of each surface slab was constrained in order to remove center of mass motion. Without a constraint on the solid, the center of mass diffusion of the entire surface will dominate the contribution to the friction kernel. Results for friction kernels were computed using individual surface atoms as adsorption sites and subsequently averaged.
    
\section{Conclusions}
\label{sec:conclusions}
    In this manuscript we have developed a theory for how surface phonons couple to molecular adsorbates based on the generalized Langevin equation. By integrating out the solid degrees of freedom (assuming they could be described harmonically) we derived a GLE for the adsorbate, wherein the friction is merely a sum of the phonon frequency of the solid weighted by their coupling to the adsorbate. 
    We demonstrated that this friction kernel depends sensitively on the frequency of the adsorbate-surface bond.
    When the frequency of this bond is smaller than the Debye frequency of the solid, adsorbates couple primarily to the acoustic phonons of solid. When the frequency of the bond is much larger than the Debye frequency of the solid, adsorbates couple primarily to the dispersionless local vibrations of the adsorption site. 
    Subsequently, we used harmonic transition state theory to derive phononic corrections to reaction rate constants. We show that these corrections improved agreement between theory and experiment for CO desorption rates from Pt(111). 

\section*{Supplementary Material}
    See supplementary material for results for friction kernels for surfaces other than Pt(111), a perturbative analysis of the physisorbed and chemisorbed limits, more details on convergence of the friction kernel with respect to supercell size, a discussion of continuum elastic theory, and more details on the derivation of rate constants using transition state theory. 
    
\section*{Data Availability}
    Data that support the findings of this study are available from the corresponding author upon reasonable request. The code used to calculate and analyze memory kernels is available on Github along with tutorials for its use \url{https://github.com/afarahva/gleqpy}.  
    
\section*{Acknowledgements}
    AF and APW were supported by the Office of Science of the U.S. Department of Energy under Contract No. DE-SC0019441.
    This research used resources of the National Energy Research Scientific Computing Center, a DOE Office of Science User Facility supported by the Office of Science of the U.S. Department of Energy under Contract No.  DE-AC02-05CH11231. 
    Ardavan Farahvash acknowledges support from the National Science Foundation Graduate Research Fellowship program. 
    
\bibliography{main.bib}

\end{document}


\title[\\]{Supplementary information for "A theory of phonon induced friction on molecular adsorbates"}

\author{Ardavan Farahvash}
\affiliation{Department of Chemistry, Massachusetts Institute of Technology, Cambridge, Massachusetts 02139, USA}

\author{Adam P. Willard}
\email{apwillard@mit.edu}
\affiliation{Department of Chemistry, Massachusetts Institute of Technology, Cambridge, Massachusetts 02139, USA}

\maketitle 

\section{Movies}

\textbf{Dynamics of acoustic flexing mode shown in the inset of Figure 1 of the main text.}

\textbf{Dynamics of local mode shown in the inset of Figure 1 of the main text.}

\section{Friction kernels and spectral densities for various solid surfaces}

    In Figures S1 through S4 we present phonon friction kernels and spectral densities for various 4x4x4 surface slabs. Figure S1 demonstrates results for a Pt(111) slab calculated modeled a Lennard-Jones (LJ) forcefield\cite{heinz_accurate_2008} and compares the results to the EMT forcefield\cite{Norskov1980} used to the main text. Figure S2 compares the results from adsorption on (111) surface facets to (100) surface facets. Figure S3. presents results for a BCC Fe(110) lattice. Figure S4. presents results for HCP Ru(0001). The Fe and Ru results were calculated using embedded atom forcefields from Ref.~\citenum{bonny_interatomic_2013} and Ref.~\citenum{fortini_asperity_2008}  respectively. Note that since Ru has a much higher Debye Frequency than Pt (225 cm$^{-1}$ for Ru and 156 cm$^{-1}$ for Pt) the dependence of the spectral density on $\omega_{\mathrm{as}}$ is shifted. All cases show the same qualitative dependence on $\omega_{\mathrm{as}}$ as presented for the Pt(111) surfaces. 
    
    \begin{figure*}[h]  
       \centering
       \includegraphics*[width=4.0in]{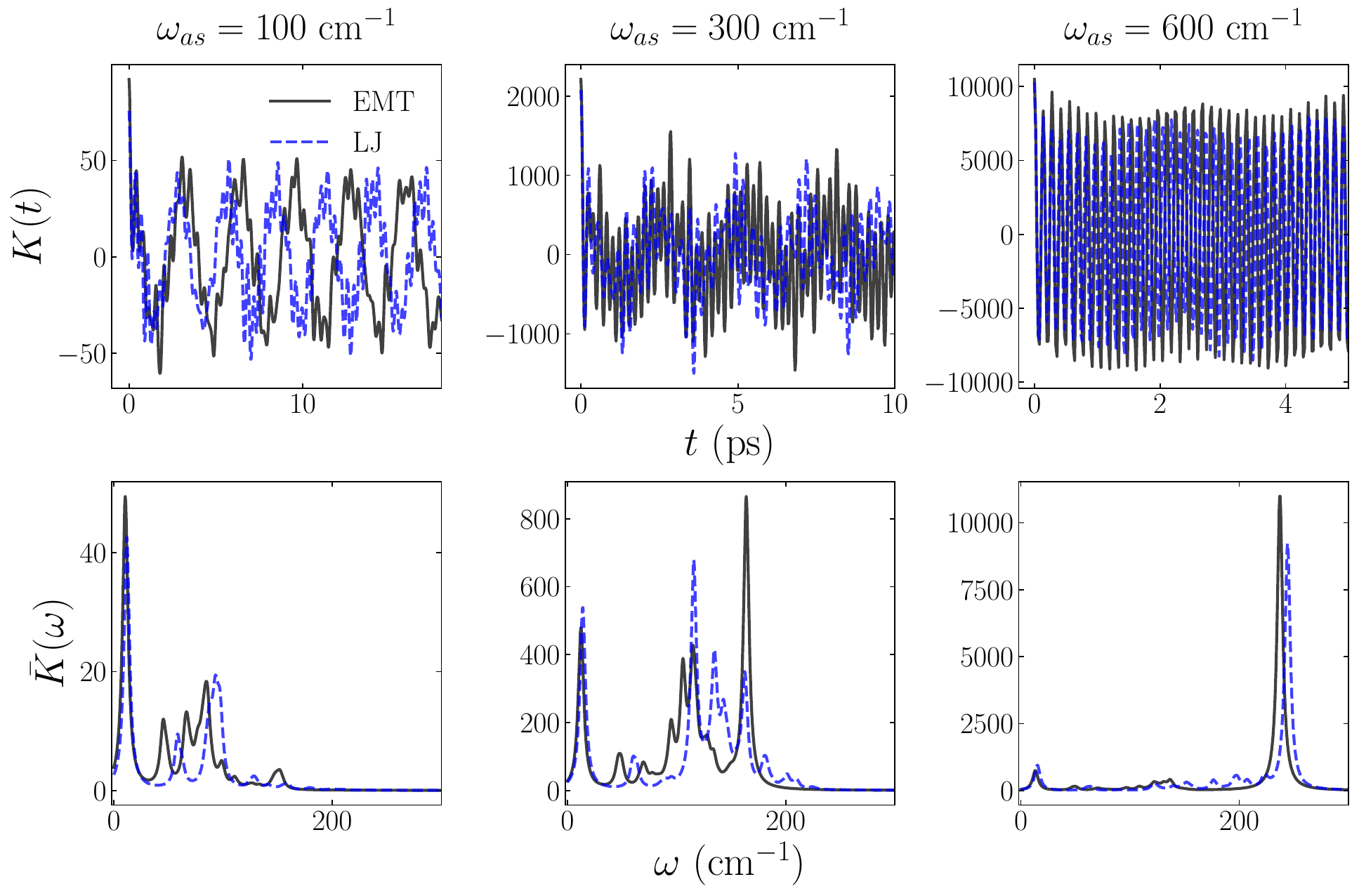}
       \caption{ (A) Friction kernel and (B) spectral density for Pt(111) surface modeled using Lennard-Jones and Effective Medium Theory forcefields.}
       \label{figure_s1}
    \end{figure*}

    \begin{figure*}[h]  
       \centering
       \includegraphics*[width=4.0in]{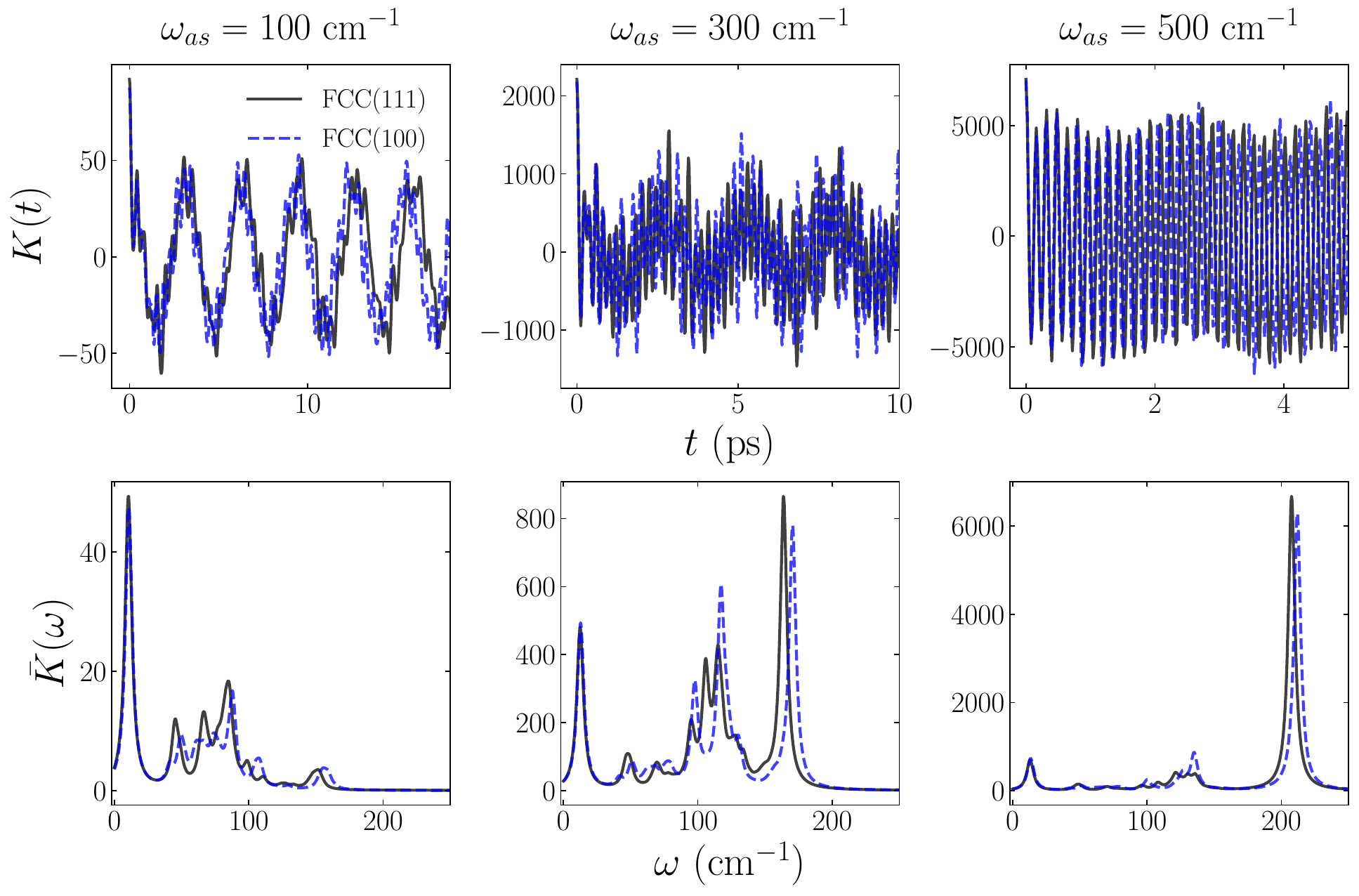}
       \caption{ (A) Friction kernel and (B) spectral density for Pt(110) surface.}
       \label{figure_s2}
    \end{figure*}

    \begin{figure*}[h]  
       \centering
       \includegraphics*[width=4.0in]{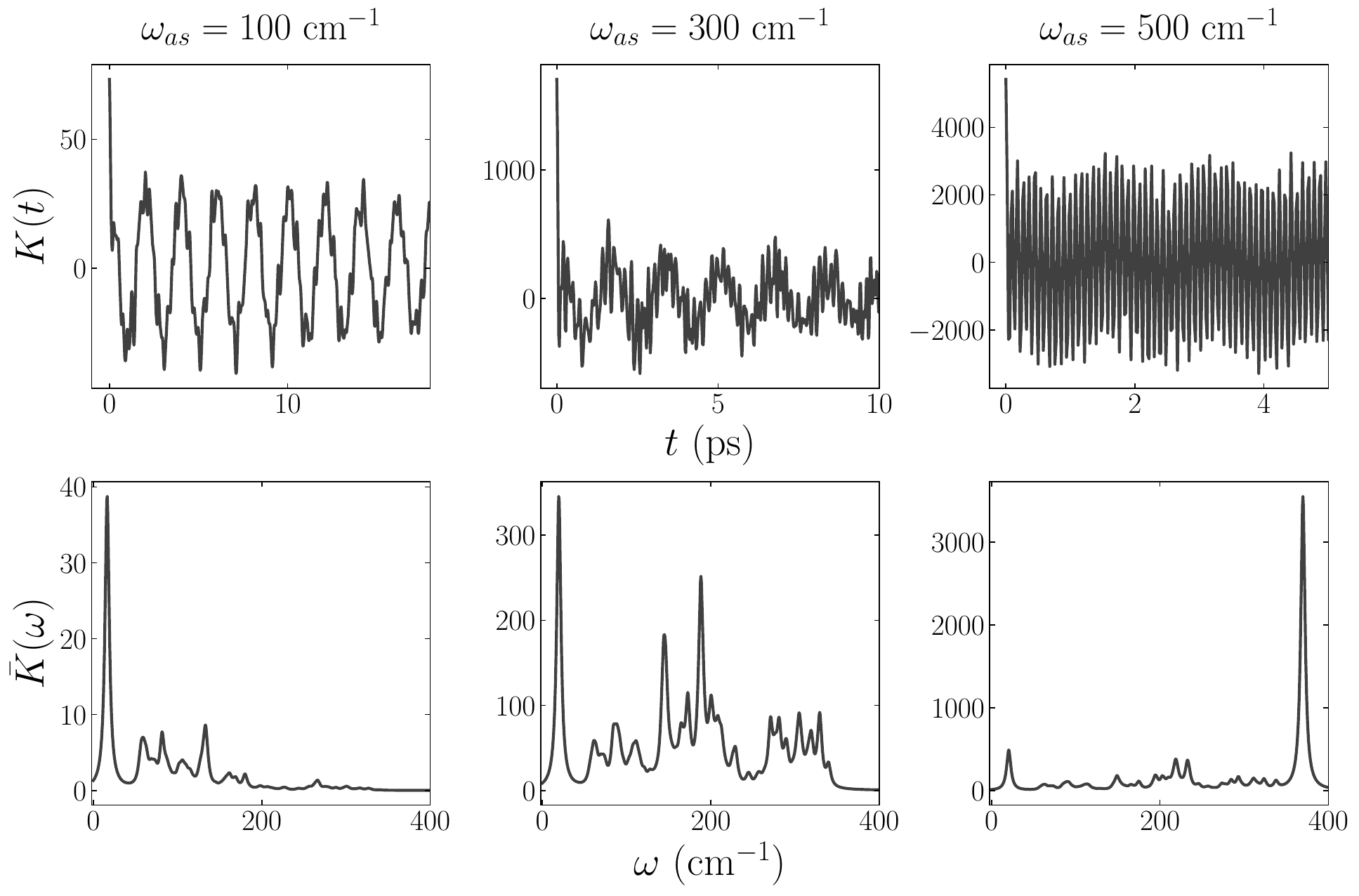}
       \caption{ (A) Friction kernel and (B) spectral density for BCC Fe(110) surface.}
       \label{figure_s3}
    \end{figure*}

    \begin{figure*}[h]  
       \centering
       \includegraphics*[width=4.0in]{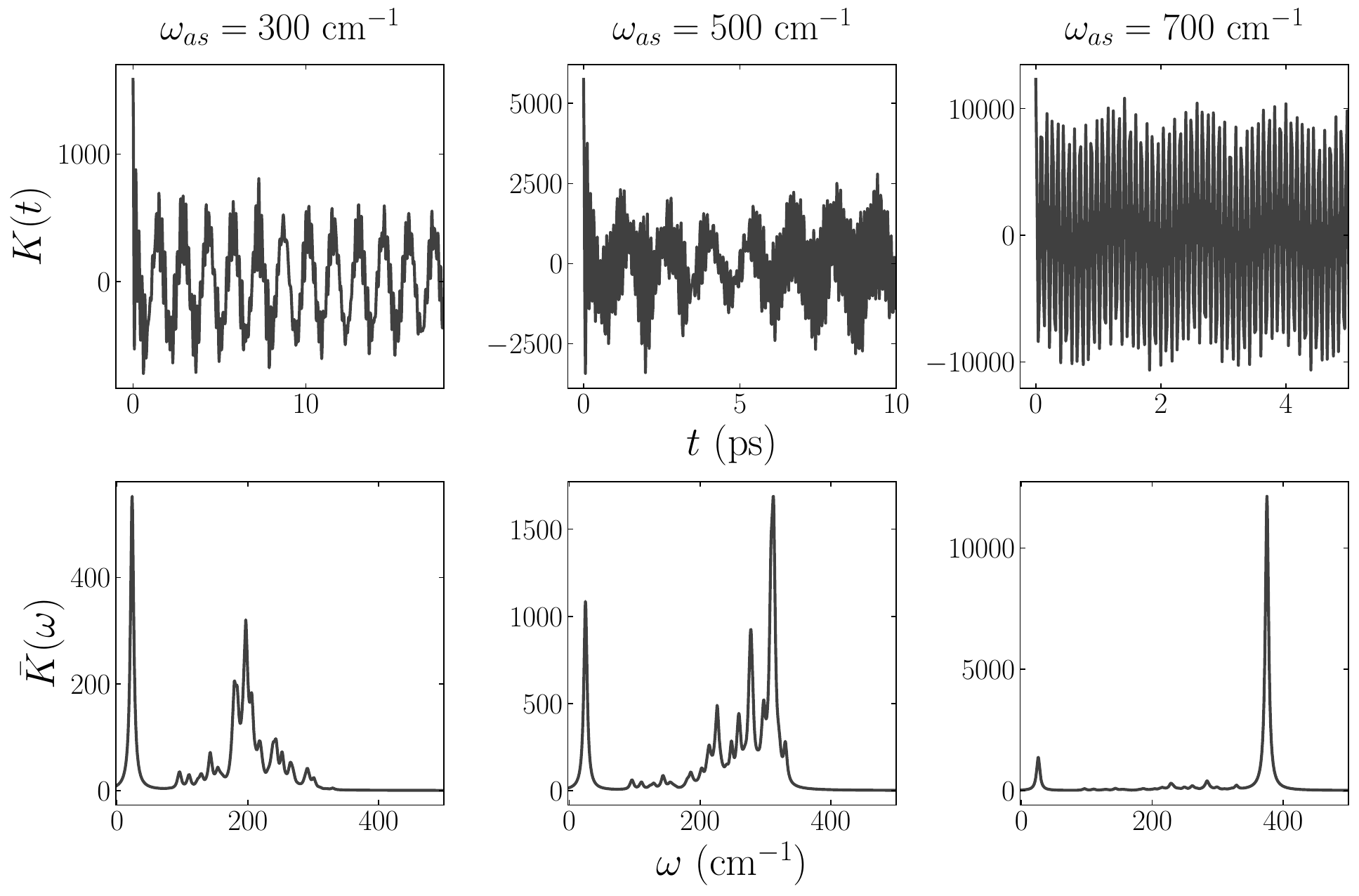}
       \caption{ (A) Friction kernel and (B) spectral density for Ru(0001) surface.}
       \label{figure_s4}
    \end{figure*}
    \FloatBarrier
    
\section{Perturbation Theory}
    In this section, we compare the results for the friction kernels and spectral densities calculated using exact diagonalization to two different perturbative schemes: one which agrees well with exact results in the chemisorbed limit (when $\omega_{\mathrm{as}}$ is large), and the other which agrees well with exact results in the physisorbed limit (when $\omega_{\mathrm{as}}$ is small).
    
    The formulas for the perturbative corrections are the familiar Rayleigh-Schrodinger perturbation theory equations. In this section, we will only use 1st and 2nd order corrections to the eigenvalues and 1st order corrections to the eigenvectors. 
    We provide the formulas below for completeness. For the eigenvalues we have,
    \begin{equation}
       \lambda_i \approx  \lambda_i^{(0)} + \lambda_i^{(1)} + \lambda_i^{(2)}
    \end{equation}
    where $\lambda_i$ is equal to the square phonon frequency $\omega^2_i$ and $\lambda_i^{(0)}$, $\lambda_i^{(1)}$, and  $\lambda_i^{(2)}$ are the $0$th, $1$st, and $2$nd order corrections respectively.
    \begin{equation}
        \lambda_i^{(1)} = \mathbf{P}^{(0)}_i \cdot \delta \mathbf{H} \cdot \mathbf{P}^{(0)}_i,
    \end{equation}
    where $\mathbf{P}^{(0)}_i$ is the $i$th eigenvector of the unperturbed Hessian and $\delta \mathbf{H} $ is the perturbation term.
    \begin{equation}
        \lambda_i^{(2)} = \sum_{j \neq i} \frac{ 
        \left( \mathbf{P}^{(0)}_j \cdot \delta \mathbf{H} \cdot \mathbf{P}^{(0)}_i \right)^2}
        {\lambda_i^{(0)} - \lambda_j^{(0)}}.
    \end{equation}
    For the eigenvectors we have,
    \begin{equation}
       \mathbf{P}_{i} \approx  \mathbf{P}^{(0)}_i + \mathbf{P}^{(1)}_i,
    \end{equation}
    where,
    \begin{equation}
       \mathbf{P}^{(1)}_i = \sum_{j \neq i} \frac{ 
        \mathbf{P}^{(0)}_j \cdot \delta \mathbf{H} \cdot \mathbf{P}^{(0)}_i }
        {\lambda_i^{(0)} - \lambda_j^{(0)}} \mathbf{P}^{(0)}_j.
    \end{equation}

    In order to gain insight from perturbation theory, we must judiciously choose how to separate the solid's mass-weighted Hessian (denoted as $\mathbf{H}_\mathrm{S}$) into the reference Hessian $\mathbf{H}_0$ and the perturbation $\delta \mathbf{H} $. In order to do so, we first structure $\mathbf{H}_\mathrm{S}$ into blocks corresponding to the surface adsorption site(s) $\mathbf{H}_\mathrm{X}$, the remaining bulk atoms $\mathbf{H}_\mathrm{Y}$, and off-diagonal blocks coupling the two $\mathbf{H}_\mathrm{XY}$,
    \begin{equation}
        \mathbf{H}_\mathrm{S} =
        \left(
        \begin{array}{c|c}
        \mathbf{H}_\mathrm{X} & \mathbf{H}_\mathrm{XY} \\ \hline
        \mathbf{H}_\mathrm{XY}^T & 
        \begin{array}{c c c} 
              &   &  \\ 
              & \mathbf{H}_\mathrm{Y} &  \\ 
              &   &  
        \end{array}
        \end{array}
        \right),
    \end{equation}
    Note that the size of $\mathbf{H}_\mathrm{Y}$ should be much larger than $\mathbf{H}_\mathrm{X}$ as most atoms in the solid can be treated as not interacting with the adsorbate. We may diagonalize $\mathbf{H}_\mathrm{Y}$ to arrive the following form,
    \begin{equation}
        \mathbf{H}_\mathrm{B} =
        \left(
        \begin{array}{c|c}
        \mathbf{H}_\mathrm{X} & \mathbf{D} \\ \hline
        \mathbf{D}^T & 
        \begin{array}{c c c} 
              &   &  \\ 
              & \pmb{\Omega}_\mathrm{Y}^2 &  \\ 
              &   &  
        \end{array}
        \end{array}
        \right),
    \end{equation}
    where $\mathbf{D}$ is the coupling between the adsorption site(s) and each bulk phonon mode, and $\pmb{\Omega}_\mathrm{Y}^2$ is a diagonal matrix containing the square frequencies of these bulk modes. With this setup, we are ready to perform perturbation theory. We will illustrate the results of perturbation theory on a 4x4x4 Pt(111) lattice with the Hessian evaluated using the EMT\cite{OpenKIM-MO:108408461881:001a} forcefield.

    \subsection{Strong coupling/chemisorbed limit}
    In the chemisorbed limit, we set the reference Hessian to one where the adsorption site(s) and the phonon modes of the bulk are uncoupled, 
    \begin{equation}
        \mathbf{H}_{0}=
        \left(
        \begin{array}{c|c}
        \mathbf{H}_\mathrm{X} & 0 \\ \hline
        0 & 
        \begin{array}{c c c} 
              &   &  \\ 
              & \pmb{\Omega}_\mathrm{Y}^2 &  \\ 
              &   &  
        \end{array}
        \end{array}
        \right),
    \end{equation}
    and therefore the perturbation is coupling,
    \begin{equation}
        \delta \mathbf{H}=
        \left(
        \begin{array}{c|c}
        0 & \mathbf{D} \\ \hline
        \mathbf{D}^T & 
        \begin{array}{c c c} 
              &   &  \\ 
              & 0 &  \\ 
              &   &  
        \end{array}
        \end{array}
        \right).
    \end{equation}
    If we assume adsorption site to be a single atom, then the following analytical forms can be found for the memory kernel to $0$th, $1$st, and $2$nd order respectively:
    \begin{equation}
        \label{eq:Kt_pert_strong0}
        K^{(0)}(t) = \frac{\mu^2}{m M} \frac{\omega_{\mathrm{as}}^4}{\tilde{\omega}_{s}^2} \cos(\tilde{\omega}_{s} t),
    \end{equation}
    \begin{equation}
        \label{eq:Kt_pert_strong1}
         K^{(1)}(t) = K^{(0)}(t) + \frac{\mu^2}{m M} \omega_{\mathrm{as}}^4 
         \sum_{i\neq s} \frac{d_i^2}{ (\tilde{\omega}^2_{s}-\omega_{\mathrm{Y},i}^2)^2 \omega_{\mathrm{Y},i}^2 } \cos(\omega_{\mathrm{Y},i} t),
    \end{equation}
    \begin{multline}
        \label{eq:Kt_pert_strong2}
         K^{(2)}(t) = \frac{\mu^2}{m M} \frac{\omega_{\mathrm{as}}^4}{\tilde{\omega}_{s}^2} 
         \cos\left( t \sqrt{\tilde{\omega}^2_{s} + \sum_{i\neq s} \frac{d_i^2}{ \tilde{\omega}^2_{s}-\omega_{\mathrm{Y},i}^2 } } \right) \\
         +
         \frac{\mu^2}{m M} \omega_{\mathrm{as}}^4 
         \sum_{i\neq s} 
         \left[
         \frac{d_i^2}{ (\tilde{\omega}^2_{s}-\omega_{\mathrm{Y},i}^2)^2 \omega_{\mathrm{Y},i}^2 + (\tilde{\omega}^2_{s}-\omega_{\mathrm{Y},i}^2)d_i^2  }
         \cos \left( t \sqrt{\omega^2_{\mathrm{Y},i} + \frac{d_i^2}{ \tilde{\omega}^2_{s}-\omega_{\mathrm{Y},i}^2 } }   \right)
         \right],
    \end{multline}
    where $d_i$ is the $i$th element $\mathbf{D}$, $\omega_{i,\mathrm{Y}}$ is $i$th element of $\pmb{\Omega}_\mathrm{Y}$, $\omega_{s}$ is the frequency of the surface adsorption site without an adsorbate bound, and $\tilde{\omega}_{s}$ is the frequency of motion of the surface adsorption site with the adsorbate bound,
    \begin{equation}
        \tilde{\omega}_{s} = \sqrt{ \frac{\mu}{M}  \omega_{\mathrm{as}}^2 + \omega_{s}^2}.
    \end{equation}

\begin{figure}[h]  
       \centering
       \includegraphics*[width=5.5in]{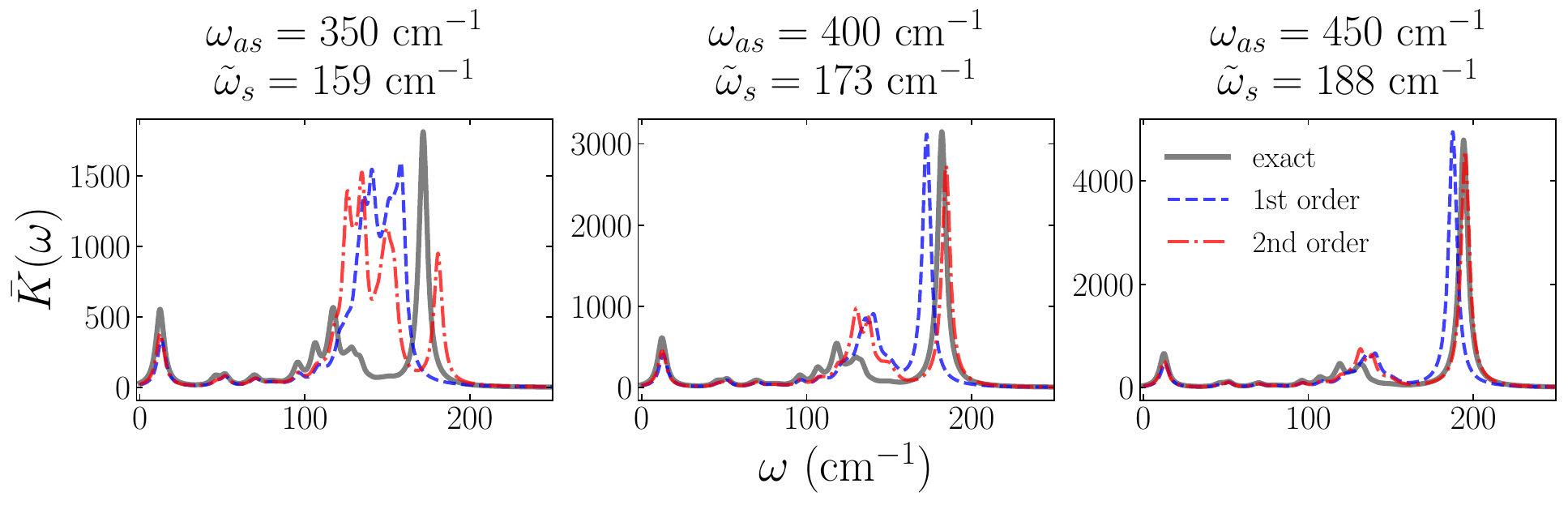}
       \caption{ Spectral density for Pt(111) surface calculated using exact diagonalization, 1st order, and 2nd order perturbation theory in the strong-coupling scheme.}
       \label{figure_s6}
    \end{figure}

    In this scheme, we see that to $0$th order the surface adsorption site does not interact with the other modes of the solid, which results in the phonon memory kernel being a single sinusoid with frequency $\bar{\omega}_\mathrm{s}$ due to solely the motion of this site. 
    The perturbation introduces coupling between the surface adsorption site and the bulk solid, allowing for the bulk phonons to contribute to the memory kernel.
    Figure~\ref{figure_s6} illustrates the spectral density calculated using this scheme to first and second order, and compares it to the results from exact diagonalization. 
    The 1st order results qualitatively match the exact results, however underestimate the frequency of the adsorption site local mode. 
    This mismatch is because the first order corrections to the eigenvalues of the Hessian are zero. 
    Introducing second order corrections removes this discrepancy, leading to excellent agreement with the exact results when $\bar{\omega}_{s} > \omega_{\mathrm{D}}$. 
    As the effective frequency of motion of the adsorption site approaches $\omega_{\mathrm{D}}$ (156 cm$^{-1}$ for Pt), this perturbative scheme qualitatively fails to describe the memory kernel/spectral density. Indeed, the results in Figure~\ref{figure_s6} show that the precise phase boundary in Figure~1 of the main text arises from the value of $\tilde{\omega}_{s}$ compared to $\omega_{\mathrm{D}}$. 
    
    \subsection{Weak-coupling/physisorbed limit}
\begin{figure}[h]  
       \centering
       \includegraphics*[width=5.5in]{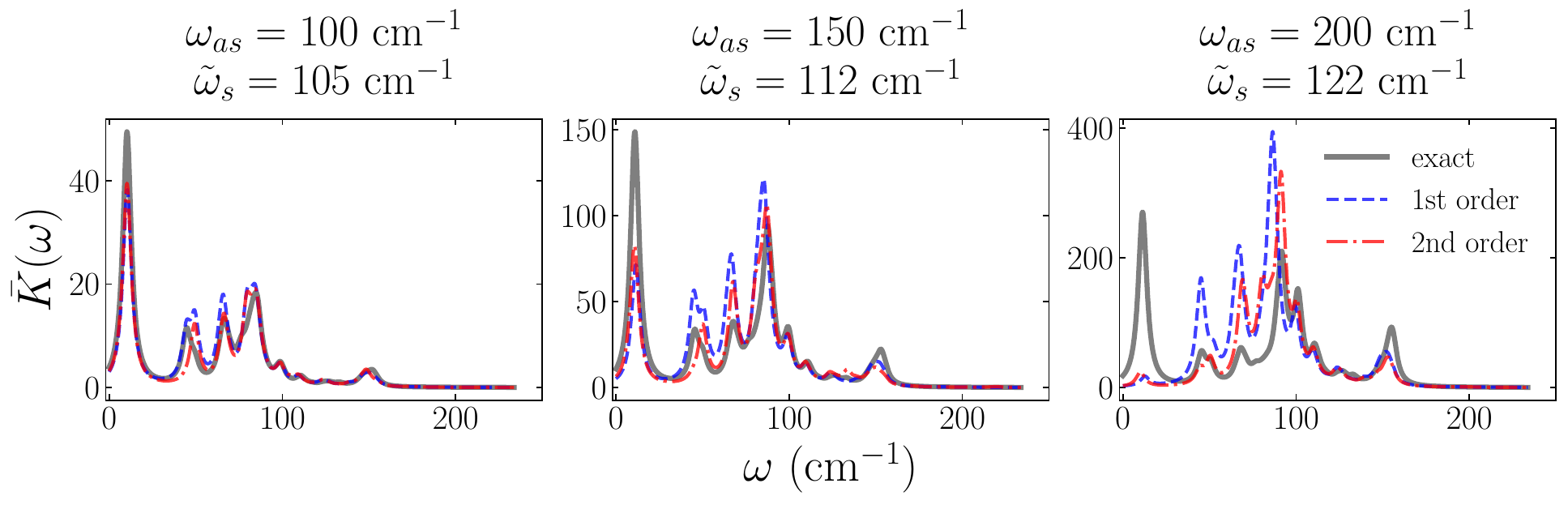}
       \caption{ Spectral density for Pt(111) surface calculated using exact diagonalization, 1st order, and 2nd order perturbation theory in the weak-coupling scheme.}
       \label{figure_s5}
    \end{figure}

    In the physisorbed limit, we treat $\mathbf{H}_\mathrm{X}$ as the perturbation, $\delta \mathbf{H} = \mathbf{H}_\mathrm{X}$, and the remaining Hessian as the reference, 
    \begin{equation}
        \mathbf{H}_{0}=
        \left(
        \begin{array}{c|c}
        0 & \mathbf{D} \\ \hline
        \mathbf{D}^T & 
        \begin{array}{c c c} 
              &   &  \\ 
              & \pmb{\Omega}_\mathrm{Y}^2 &  \\ 
              &   &  
        \end{array}
        \end{array}
        \right).
    \end{equation}
    This scheme assumes that the contribution of the adsorption sites to the bulk phonon modes is small. 
    Figure~\ref{figure_s5} illustrates the spectral density calculated using this scheme, to first and second order, and compares it to the results from exact diagonalization. 
    The 1st order results for $K(\omega)$ use 1st order corrections for both the eigenvectors and eigenvalues, while the 2nd order results add 2nd order corrections to the eigenvalues while keeping the eigenvectors at 1st order.  
    Both 1st and 2nd order results agree well with the exact results when $\omega_{\mathrm{as}}$ is less than platinum's Debye frequency $\omega_{\mathrm{D}}=156$ cm$^{-1}$ as expected, and even qualitatively capture results at $\omega_{\mathrm{as}} \approx \omega_{\mathrm{D}}$. 
    However, once again we see that as the effective frequency of motion of the adsorption site ($\omega_s$) approaches $\omega_{\mathrm{D}}$, the perturbative scheme fails. 
    \FloatBarrier
    
\newpage
\section{Phonon Dispersion}
\begin{figure}[h]  
       \centering
       \includegraphics*[width=5.5in]{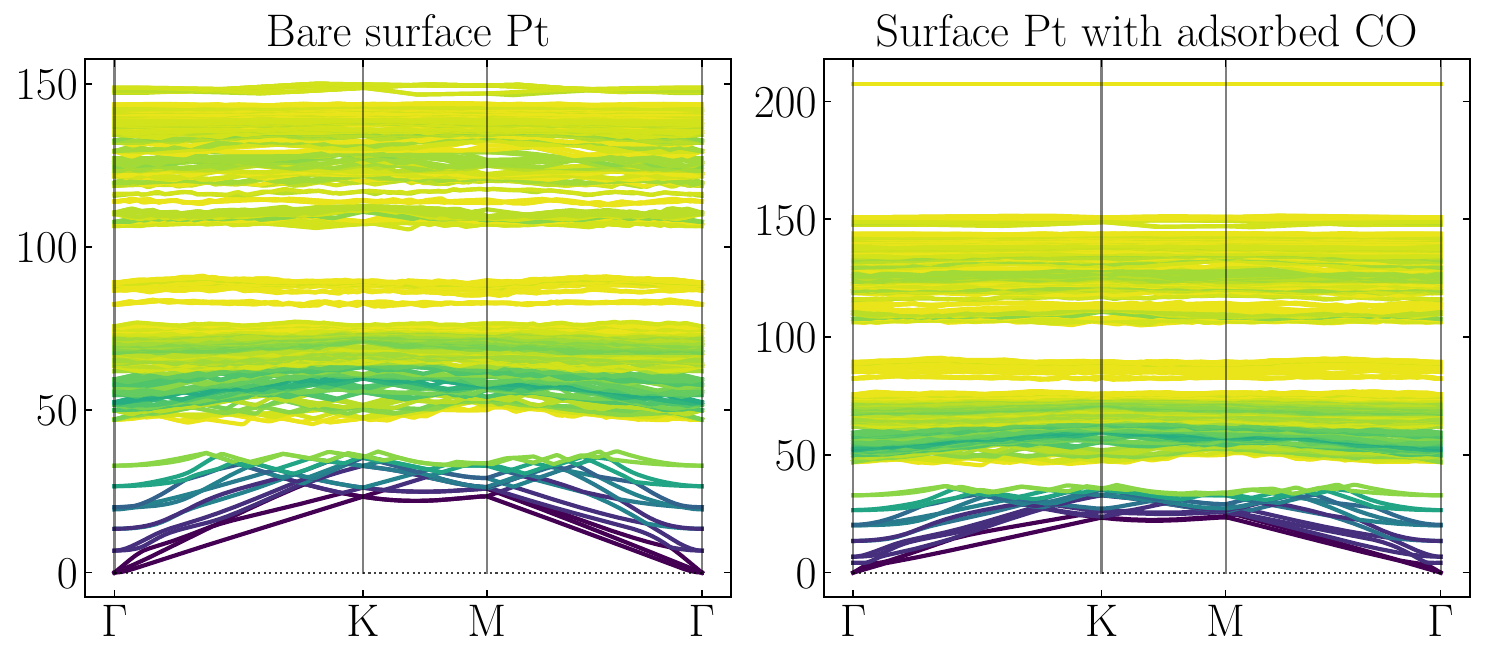}
       \caption{ Phonon dispersion curves for a 4x4x16 EMT Pt(111) surface slab.}
       \label{figure_s7}
    \end{figure}

\begin{figure}[h]  
       \centering
       \includegraphics*[width=5.5in]{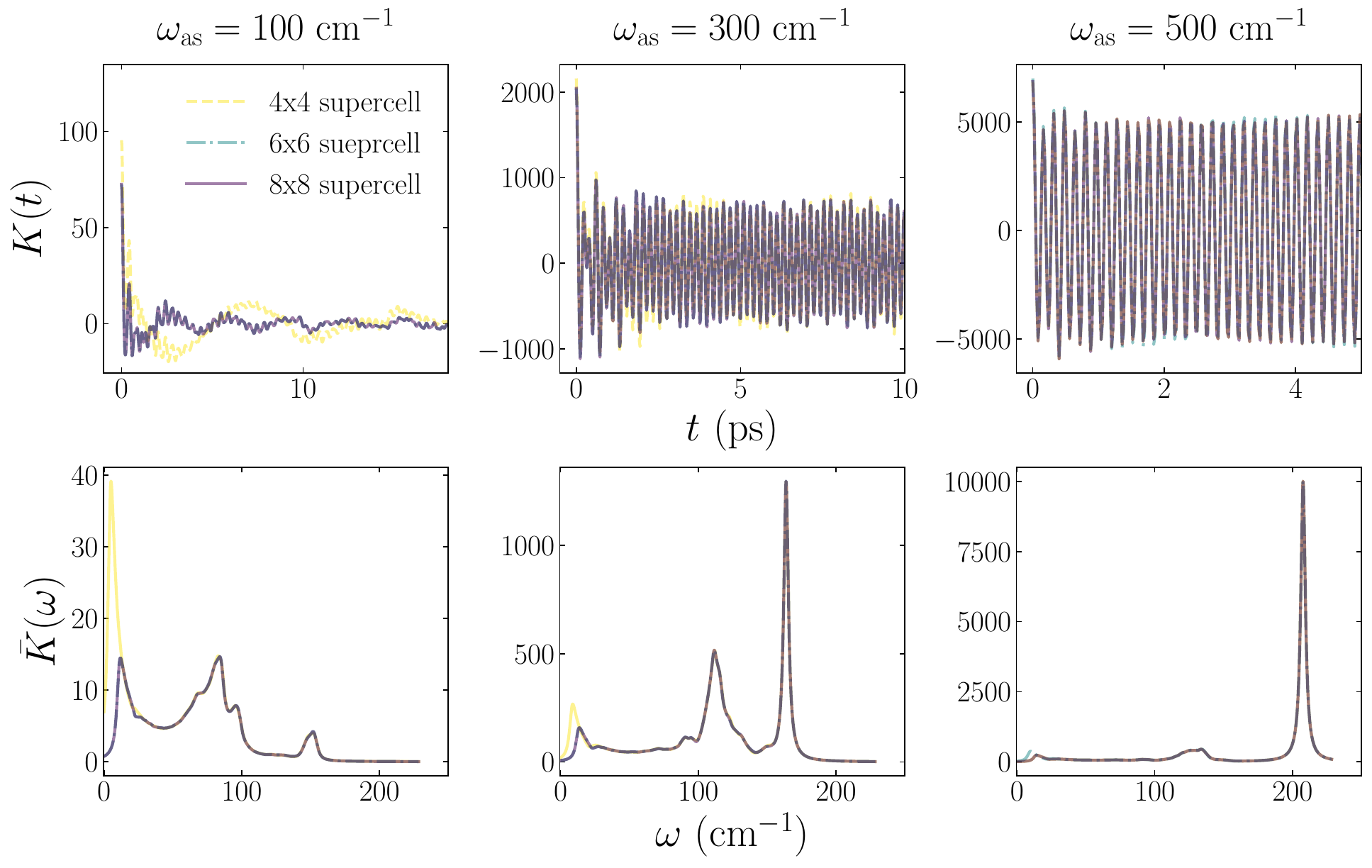}
       \caption{ Phonon friction kernels and spectral densities averaged over $k$-space for different values of the number of the size of the surface super-cell used in computing the dynamical matrix.}
       \label{figure_s8}
    \end{figure}

    In Fig.~\ref{figure_s7} we present the surface phonon dispersion curves for a 4x4x16 Pt(111) surface slab. In Fig.~\ref{figure_s8} we present the convergence of the phonon spectral density as a function of the surface cell used to calculate the phonon frequencies. 
    
    \FloatBarrier

\section{Elastic Continuum Theory}
    In this section, we take an alternative approach to determining the phonon-induced friction in the limit of a macroscopic solid by using continuum elastic theory instead of atomistic models. 
    Such an approach was originally explored by Ref.~\citenum{diebold_generalized_1979} and Ref.~\citenum{persson_brownian_1985}. It is theoretically appealing due to the minimal, experimentally accessible free parameters used in elastic theory. However, it also suffers from several limiting assumptions which we explicitly delineate.
    
    Elastic (energy conserving) acoustic waves in a material may be modeled via the Navier-Cauchy equation,
    \begin{equation}
        \label{eq:navier_cauchy}
        \ddt{\mathbf{u}}(\mathbf{r},t) = c_t \vec{\nabla}^2 \mathbf{u}(\mathbf{r},t) 
        + (c_l^2 + c_t^2) \nabla \left( \nabla \cdot \mathbf{u}(\mathbf{r},t)\right)
        + \mathbf{F}(\mathbf{r},t),
    \end{equation}
    where $\mathbf{u}(\mathbf{r},t)$ is the displacement of the solid at position $\mathbf{r}=(x,y,z)$ and time $t$, $\vec{\nabla}^2$ is the vector Laplacian, $\nabla \left( \nabla \cdot \right)$ is gradient of the divergence, and $\mathbf{F}$ are external forces. Solutions to this Eq.~\ref{eq:navier_cauchy} can generally be separated into zero divergence and zero curl components corresponding to the transverse and longitudinal modes respectively,
    \begin{equation}
        \label{eq:seperation_of_modes}
        \ddt{\mathbf{u}}(\mathbf{x},t) = \mathbf{u}_l(\mathbf{x},t) + \mathbf{u}_t(\mathbf{x},t),
    \end{equation}
    each of which satisfy a 3D wave equation,
    \begin{equation}
        \label{eq:wave_equation}
        \ddt{\mathbf{u}}_{l/t}(\mathbf{x},t) = c_{l/t} \vec{\nabla}^2 \mathbf{u}_{l/t}(\mathbf{x},t).
    \end{equation}

    Consider a single adsorbate degree of freedom whose displacement is denoted by $q$. If this degree of freedom is harmonically coupled to surface normal (z-axis) displacement the solid at position $\mathbf{r}_0=(x_0,y_0,L_z)$ then the coupled adsorbate-solid equations are,
    \begin{equation}
        \label{eq:ads_ode}
        \ddot{q}(t) = -\frac{1}{m} \frac{d V_A}{d q}(t) - \frac{\mu}{m} \omega_{as}^2(q(t) - u_z(\mathbf{r}_0,t))
    \end{equation}
    \begin{equation}
        \label{eq:solid_pde}
        \ddt{\mathbf{u}}(\mathbf{r},t)
        - c_t \vec{\nabla}^2 \mathbf{u}(\mathbf{r},t) 
        - (c_l^2 + c_t^2)\nabla \left( \nabla \cdot \mathbf{u}(\mathbf{r},t)\right)
        = 
        \frac{\mu \omega_{as}^2}{M} (q(t)-u_z(\mathbf{r}_0,t))  a^3 \delta \left( \mathbf{r} - \mathbf{r}_0 \right) \hat{z},
    \end{equation}
    where $\hat{z}=(0,0,1)$ is the unit vector in the surface normal, $a$ is spacing between atoms in the crystal, and $M$ is the mass of the solid atom. The $a^3$ factor arises from taking the continuum limit of a force on a single lattice point and offsets the inverse volume units of the 3D delta function $\delta \left( \mathbf{r} - \mathbf{r}_0 \right)$. The forces from the adsorbate on the solid can be be separated in two contributions. A static contribution,
    \begin{equation}
        \label{eq:fads_static}
        \frac{\mu \omega_{as}^2}{M} u_z(\mathbf{r}_0,t)  a^3 \delta \left( \mathbf{r} - \mathbf{r}_0 \right),
    \end{equation}
    which enforces to the shift in the solid's vibrational spectrum due to the presence of the adsorbate, and a dynamic contribution,
    \begin{equation}
        \label{eq:fads_time}
        f(\mathbf{r},t) = \frac{\mu \omega_{as}^2}{M} q(t)  a^3 \delta \left( \mathbf{r} - \mathbf{r}_0 \right),
    \end{equation}
    representing the time-dependent external force of the adsorbate on the solid. 
    We can rearrange Eq.~\ref{eq:solid_pde} as,
    \begin{equation}
        \label{eq:solid_pde2}
        \left[ \frac{d^2}{dt^2} - c_t^2 \vec{\nabla}^2 - (c_l^2 + c_t^2) \nabla \nabla \cdot + \frac{\mu \omega_{as}^2}{M} a^3 \delta(\mathbf{r} - \mathbf{r}_0) \hat{z} \right] \mathbf{u}(\mathbf{r},t) = f(\mathbf{r},t)\hat{z}.
    \end{equation}
    The operator on the right-hand-side (RHS) of this equation is a linear operator; therefore Eq.~\ref{eq:solid_pde2} may be solved using the method of Green's functions,
    \begin{equation}
        \label{ref:greens_solution}
        \mathbf{u}(\mathbf{r},t) = \mathbf{u}_0(\mathbf{r},t) + 
        \int_0^t d\tau \int d\mathbf{r}' \mathbf{G}(\mathbf{r},t;\mathbf{r}',\tau) \cdot f(\mathbf{r}',\tau) \hat{z}
    \end{equation}
    where $\mathbf{u}_0(x,t)$ is the solution to the homogeneous case and $\mathbf{G}(\mathbf{r},t;\mathbf{r}',\tau)$ is a 3x3 tensor Green's function corresponding to the operator on the RHS of Eq.~\ref{eq:solid_pde2}. If we substitute this solution into Eq.~\ref{eq:ads_ode} we arrive at a GLE where the memory kernel is proportional to the antiderivative of the Green's function,
    \begin{equation}
        \label{eq:memory_greens}
        K(t) = \frac{\mu^2 \omega^4_{as}}{m M} a^3 \int dt G_{zz}(\mathbf{r}_0,t;\mathbf{r}_0,0).
    \end{equation}
    Henceforth we will denote $G_{zz}(\mathbf{r}_0,t;\mathbf{r}_0,0)$ as simply $G(t)$ for simplicity.
    This Green's function may be decomposed in the following form,
    \begin{equation}
        \label{eq:greens_sum}
        G(t) = \sum_{\alpha} \sum_{\mathbf{k}}
        \frac{ \sin(c_{\alpha} \left| \mathbf{k} \right| t)}{c_{\alpha} \left| \mathbf{k} \right|}
        R_{z,\alpha}(\mathbf{r}_0,\mathbf{k})  R_{z,\alpha}^*(\mathbf{r}_0,\mathbf{k}) 
    \end{equation}
    where $\alpha$ denotes phonon polarizations (i.e. transverse or longitudinal), and $R_{z,\alpha}$ are the $z$th spatial components of the normalized eigenfunctions of the operator on the RHS of Eq.~\ref{eq:solid_pde2}. The spectrum of $\mathbf{k}$ values as well as the specific form of the spatial eigenfunctions depend on the choice of boundary conditions. 

    Due to the delta function in the adsorbate shift term (Eq.~\ref{eq:fads_static}), the allowed $\mathbf{k}$ values and cannot be computed exactly. Indeed, this term makes Eq.~\ref{eq:solid_pde2} very similar to the Schr\"{o}dinger Equation with a delta function well, in which the spectrum must be computed numerically as a solution to a system of transcendental equations. However, perturbation theory, physical intuition, and the numerical results presented in Figure~2 of the main text all suggest that the low-frequency acoustic modes of a solid should not depend on the presence of an adsorbate. Therefore, we proceed by ignoring the adsorbate shift term while noting that, by construction, such an approach is only valid for the low-frequency acoustic modes.
    Setting periodic boundary conditions in the $xy$ plane, fixed boundary conditions at $z=0$ and Neumann boundary conditions at $z=L_z$ we have,
    \begin{equation}
        \label{eq:modes_pbc}
        R_{z,\alpha}^*(\mathbf{r},\mathbf{k}) = \frac{2}{\sqrt{L_x L_y L_z}} e^{2 \pi i k_x x} e^{2 \pi i k_y y} \sin(k_z z)
    \end{equation}
    where $L_x$, $L_y$, and $L_z$ are the size of the solid in the $x$, $y$, and $z$ directions respectively. The allowed are values of $k$ are $k_{x} = \frac{2 \pi n_x}{L_x}$, $k_{y} = \frac{2 \pi n_y}{L_y}$, and $k_{z} = \frac{(n_z+\frac12)\pi }{L_z}$, where $n_x$ and $n_y$ are any integers and $n_z$ is any integer greater than or equal to zero. 
    Taking the limit as $L_x$, $L_y$, and $L_z$ become very large we find that the Green's function becomes,
    \begin{equation}
        \label{eq:Gt_infinite_1}
        G(t) = \sum_{\alpha} \frac{1}{8 \pi^3} \int d\mathbf{k} 
        \frac{ \sin(c_{\alpha} \left| \mathbf{k} \right| t)}{c_{\alpha} \left| \mathbf{k} \right|}.
    \end{equation}
    It is well-known that the integral in Eq.~\ref{eq:Gt_infinite_1} diverges if the integral is taken over all $k$-space due to the contribution wavelengths smaller than the inter-atom spacing. Therefore, the integral in Eq.~\ref{eq:Gt_infinite_1} should only be taken over first Brillouin zone. Taking inspiration from the Debye model, we may approximate the first Brillouin zone with a radial cutoff $\left| k_{\mathrm{D}} \right|$,
    \begin{equation}
        \label{eq:Gt_infinite_2}
        G(t) = \frac{1}{2 \pi^2} \sum_{\alpha}  \int_0^{k_\mathrm{D}} dk
        \frac{ \sin(c_{\alpha} k t)}{c_{\alpha} {k}}.
    \end{equation}
    Carrying out the integration over $k$ and subsequently integrating over time $t$, leads to the following formulas for the memory kernel and spectral density, 
    \begin{equation}
        \label{eq:memory_elastic_final}
        K_{\mathrm{cont}}(t) = \frac{\mu^2 \omega^4_{\mathrm{as}}}{m M} \frac{a^3}{2 \pi^2} \left( \frac{2}{c_t^3} + \frac{1}{c_l^3} \right) \frac{\sin\left(\omega_{\mathrm{D}} t\right)}{t}.
    \end{equation}
    \begin{equation}
        \label{eq:spectral_elastic_final}
        \bar{K}_{\mathrm{cont}(\omega)} = \frac{\mu^2 \omega^4_{\mathrm{as}}}{m M} \frac{a^3}{2 \pi^2} \left( \frac{2}{c_t^3} + \frac{1}{c_l^3} \right) \Theta(\omega-\omega_{\mathrm{D}}).
    \end{equation}
    where $\Theta$ is the Heaviside step function. Eq.~\ref{eq:spectral_elastic_final} illustrates that the memory is flat (Ohmic) with a high frequency cutoff at the Debye frequency. 

    \begin{figure}[h]  
       \centering
       \includegraphics*[width=3.25in]{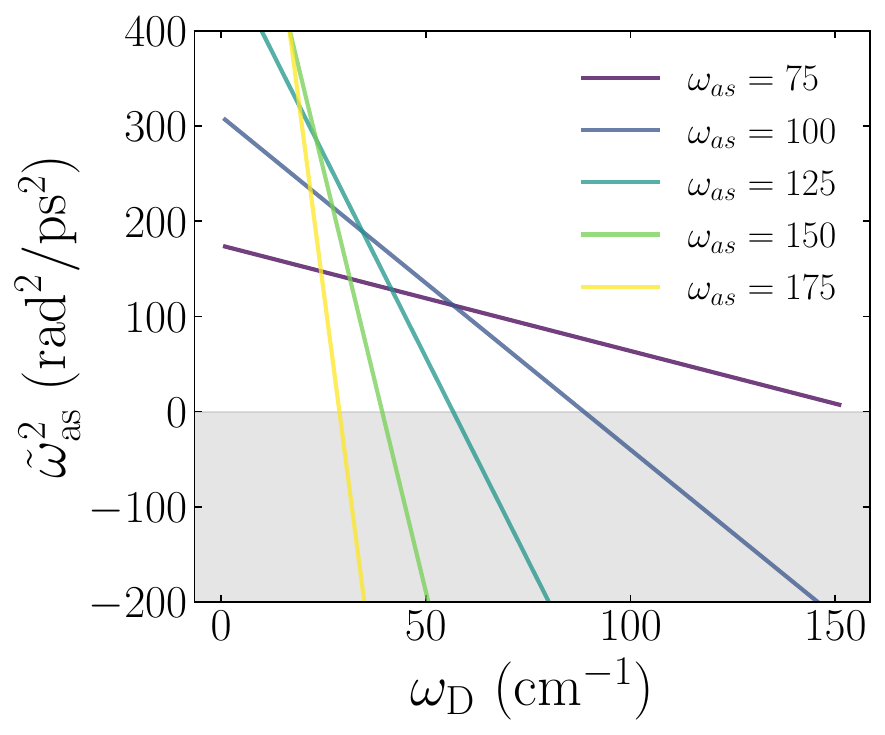}
       \caption{Square of the effective (phonon-renormalized) surface-adsorbate bond frequency as a function of the solid's Debye frequency.}
       \label{figure_s9}
    \end{figure}
    We can use Eq~\ref{eq:memory_elastic_final} to compute the effective adsorbate-surface bond frequency,
    \begin{equation}
        \label{eq:ads_surf_bond_suppl}
        \tilde{\omega}_{\mathrm{as}} = \sqrt{\frac{\mu}{m}\omega^2_{\mathrm{as}} - K(t=0)}.
    \end{equation}
    Figure~\ref{figure_s9} illustrates results for $\tilde{\omega}_{\mathrm{as}}^2$ as a function of the Debye frequency and ${\omega}_{\mathrm{as}}$. One can see that for physically reasonable choices of parameters (including those for CO adsorbed on Pt) the effective frequency becomes imaginary, signifying that there is no stable adsorption state. Such a clearly unphysical conclusion is a result of the ignoring the adsorbate shift term (Eq.~\ref{eq:fads_static}) in the original equations of motion and its concomitant effects on the Green's function/memory kernel. Thus, we again emphasize that the functional forms in Eq.~\ref{eq:memory_elastic_final} and Eq.~\ref{eq:spectral_elastic_final} are only valid for the low-frequency acoustic modes of the solid, and generally integrating all $\mathbf{k}$ vectors up to the Debye frequency is inappropriate and a lower frequency cutoff should be used. 
    \FloatBarrier

\section{Reaction rate theory}
    \subsection{Transition state theory}
    Transition state theory calculates the rate constant of a chemical reaction as the equilibrium flux of trajectories through a dividing surface separating reactants and products. Formulaically, such a rate constant may be expressed as, 
    \begin{equation}
        \label{eq:k_tst}
        k = \frac{1}{Q_R} \int d\mathbf{r} d\mathbf{p} \
        e^{-\beta H(\mathbf{r},\mathbf{p})} \delta\left[ f(\mathbf{r}) \right] 
        \left( \nabla f \cdot \mathbf{p} \right) \Theta \left( \nabla f \cdot \mathbf{p} \right),
    \end{equation}
    where $\mathbf{r}$ and $\mathbf{p}$ are mass-weighted positions and momenta respectively, $H$ is the Hamiltonian,
    $f(\mathbf{x})=0$ is the dividing surface, $\nabla f$ is the surface normal, $\Theta$ is the Heaviside step function, and $Q_R$ is the reactant partition function. $Q_R$ is defined as,
    \begin{equation}
        \label{eq:Q_R}
        Q_0= \int_0 d\mathbf{r} d\mathbf{p} \ e^{-\beta H(\mathbf{r},\mathbf{p})}.
    \end{equation}
    where the subscript $0$ denotes that the integral is taken only over the reactant region in position space. The transition state is the saddle point on this dividing surface and, in the vicinity of the transition state, the reaction coordinate corresponds to the unstable (imaginary frequency) normal mode.
    Carrying out the momentum integrals in Eq.~\ref{eq:k_tst} (assuming a one-dimensional reaction coordinate) we have,
    \begin{equation}
        \label{eq:k_tst2}
        k = \frac{1}{\sqrt{2 \pi \beta}} \frac{
        \int d\mathbf{r} \ e^{-\beta V(\mathbf{r})} \delta\left[ f(\mathbf{r}) \right],
        }
        {
        \int_R d\mathbf{r} \ e^{-\beta V(\mathbf{r})}.
        }
    \end{equation}
    The integral in the numerator is largest in the vicinity of the transition state while the integral in the denominator is largest near the minimum of $V$. Expanding around these two points gives,
    \begin{equation}
        \label{eq:V_min}
        V(\mathbf{r} \approx \mathbf{r}_0) 
        = E_R - \frac12 (\mathbf{r}-\mathbf{r}_0)^T \mathbf{H}^{(0)} (\mathbf{r}-\mathbf{r}_0),
    \end{equation}
    for the reactant basin and,
    \begin{equation}
        \label{eq:V_ts}
        V(\mathbf{r} \approx \mathbf{r}^{\ddag}) 
        = E^{\ddag} - \frac12 (\mathbf{r}-\mathbf{r}^{\ddag})^T \mathbf{H}^{\ddag} (\mathbf{r}-\mathbf{r}^{\ddag}),
    \end{equation}
    for the transition state, where $\mathbf{H}$ is the mass-weighted Hessian evaluated at the minimum of the reactant basin and $\mathbf{H}^{\ddag}$ is the mass-weighted Hessian evaluated at the transition state. Using Eq.~\ref{eq:V_min} and Eq.~\ref{eq:V_ts} we may evaluate the integrals in Eq.~\ref{eq:k_tst2} giving, 
    \begin{equation}
        \label{eq:k_tst3}
        k = \frac{\lambda^{\ddag}}{2 \pi} 
        \sqrt{ 
        \frac{\textrm{det}(\mathbf{H}^{(0)}}{-\textrm{det}(\mathbf{H}^{\ddag})}
        }
        e^{-\beta(E^{\ddag}-E_R)},
    \end{equation}
    where $\lambda^{\ddag}$ is the norm of the frequency of the unstable mode of $\mathbf{H}^{\ddag}$, and "$\textrm{det}$" denotes the matrix determinant. 
    For a reaction at a surface, these Hessians can be organized into a block structure corresponding to the molecular/adsorbate degrees of freedom, the solid degrees of freedom, and the coupling between them,
    \begin{equation}
        \label{eq:block_hessian_TST}
        \mathbf{H} =
        \left(
        \begin{array}{c|c}
        \mathbf{H}_{A} & \mathbf{G}_{AS} \\ \hline
        \mathbf{G}_{AS}^T & 
        \begin{array}{c c c} 
              &   &  \\ 
              & \mathbf{H}_{S} &  \\ 
              &   &  
        \end{array}
        \end{array}
        \right).
    \end{equation}
    The determinant of such a block matrix may be evaluated as, 
    \begin{equation}
        \label{eq:determinant_identity}
        \textrm{det}(\mathbf{H}) = \textrm{det}(\mathbf{H}_{S}) \times 
        \textrm{det}(\mathbf{H}_{A} -  \mathbf{G}_{AS}^T \mathbf{H}^{-1}_{S}\mathbf{G}_{AS}).
    \end{equation}
    Using this determinant identity together with Eq.~\ref{eq:k_tst3} gives Eq.~25 of the main text. 
    
    \subsection{Kramers-Grote-Hynes theory}
    The effect of friction on reaction rates is explicitly captured in Kramers-Grote-Hynes (KGH) theory, wherein the rate constant is,
    \begin{equation}
        \label{eq:k_kgh}
        k_\mathrm{KGH} = \frac{\tilde{f}_0}{2 \pi} \frac{\lambda^\ddag}{\tilde{f}^\ddag} e^{-\beta \Delta E^{\ddag} }, 
    \end{equation}
    where $\tilde{f}_0$ is the effective (thermodynamically renormalized) frequency of the reactant basin, $\tilde{f}^\ddag$ is the effective barrier frequency, and $\lambda^\ddag$ is related to the friction kernel by,
    \begin{equation}
        \label{eq:lam_kgh}
        -(\lambda^\ddag)^2 + \lambda^\ddag \bar{K}(\lambda=\lambda^\ddag) + (f^{\ddag})^2 = 0,
    \end{equation}
    and $\bar{K}(\lambda)$ is the Laplace transform of the friction kerrnel. Given that the friction kernel arises explicitly in KGH theory it is appealing to directly apply Eq.~\ref{eq:k_kgh} to evaluate the effect of phonons on surface reaction rates. However, doing so only corresponds to using a more limited form Eq.~\ref{eq:k_tst3}. Below we demonstrate how to derive Eq.~\ref{eq:k_kgh} from Eq.~\ref{eq:k_tst3}, noting that that such a relationship was originally established by Pollak\cite{pollak_theory_1986}.

    We begin by assuming that the reaction coordinate is one-dimensional and ignoring other molecular degrees of freedom besides this reaction coordinate. Under such assumptions $\mathbf{H}_{A}$ has only one entry. For the reactant basin this entry is the square frequency of the well $\mathbf{H}^{(0)}_\mathrm{A} = {f}^2_0$ and for the transition-state this entry is the square frequency of the barrier $\mathbf{H}^{(\ddag)}_\mathrm{A} = -({f}^\ddag)^2$. The Hessian for the reactant basin is now, 
    \begin{equation}
        \label{eq:hessian_react}
        \mathbf{H}^{(0)} =
        \left(
        \begin{array}{c|c}
        f^2_0 & \mathbf{G}^{(0)}_{AS} \\ \hline
        \mathbf{G}^{(0)}_{SA} & 
        \begin{array}{c c c} 
              &   &  \\ 
              & \mathbf{H}^{(0)}_{S} &  \\ 
              &   &  
        \end{array}
        \end{array}
        \right), 
    \end{equation}
    and the Hessian for the transition state is, 
    \begin{equation}
        \label{eq:hessian_TS}
        \mathbf{H}^{\ddag} =
        \left(
        \begin{array}{c|c}
        -(f^\ddag)^2 & \mathbf{G}^{\ddag}_{AS} \\ \hline
        \mathbf{G}^{\ddag}_{SA} & 
        \begin{array}{c c c} 
              &   &  \\ 
              & \mathbf{H}^{\ddag}_{S} &  \\ 
              &   &  
        \end{array}
        \end{array}
        \right). 
    \end{equation}
    We must also assume that the solid phonon modes and the adsorbate-solid coupling are equal in both reactant basin and transition well; that is $\mathbf{G}^{\ddag}_\mathrm{AS} = \mathbf{G}^{(0)}_\mathrm{AS}$ and $\mathbf{H}^{\ddag}_\mathrm{S} = \mathbf{H}^{(0)}_\mathrm{S}$. Many surface-chemical processes may violate such assumptions. For example, in a desorption process the reactant basin may have a very strong coupling to the phonon modes, while the transition state usually does not. However, in KGH theory it is required that the friction kernel, and by extension $\mathbf{H}_\mathrm{S}$ and  $\mathbf{G}_\mathrm{AS}$, remain the same at every point along the reaction coordinate.

    These assumptions simplify the Hessian determinants significantly. In particular, for the reactant determinant we have,
    \begin{equation}
        \textrm{det}(\mathbf{H}^{(0)}) = \textrm{det}(\mathbf{H}_S) \cdot \left( {f}^2_0 - \mathbf{G}_{AS}^T \mathbf{H}^{-1}_{S} \mathbf{G}_{AS} \right)
    \end{equation}
    and for the transition state determinant we have, 
    \begin{equation}
        \textrm{det}(\mathbf{H}^{\ddag}) = \textrm{det}(\mathbf{H}_S) \cdot \left( -({f}^\ddag)^2 - \mathbf{G}_{AS}^T \mathbf{H}^{-1}_{S} \mathbf{G}_{AS} \right).
    \end{equation}
    With these determinants, and noting that $K(t=0) = \mathbf{G}_{AS}^T \mathbf{H}^{-1}_{S} \mathbf{G}_{AS}$, we can simplify Eq.~\ref{eq:k_tst3} to,
    \begin{equation}
        k = \frac{\tilde{f}_0}{2 \pi} \frac{\lambda^\ddag}{\tilde{f}^\ddag} e^{-\beta \Delta E^{\ddag} }, 
    \end{equation}
    where $\tilde{f}_0 = \sqrt{ {f}_0^2 - K(t=0) }$ and $\tilde{f}^\ddag = \sqrt{ (\tilde{f}^\ddag)^2 + K(t=0) }$. 
    
    The remaining step is to show that $\lambda^\ddag$ satisfies Eq.~\ref{eq:lam_kgh}. To do so, first note that the Laplace transform of the friction kernel may be expressed as,
    \begin{equation}
        \label{eq:mem_laplace}
        \bar{K}(\lambda) = \int_0^\infty d\omega \bar{K}(\omega) \frac{\lambda}{\lambda^2 + \omega^2}
    \end{equation}
    where $\bar{K}(\omega)$ is the spectral density defined in Eq.~13 of the main text. The equation $\textrm{det}(\mathbf{H}^{(\ddag)} - \lambda^2 \mathbf{I}) = 0$ sets the eigenfrequencies in the vicinity of the barrier including $\lambda^\ddag$. Evaluating this determinant leads to,
    \begin{equation}
        \label{eq:barrier_determinant}
        \left[ \prod_{i=1}^N (\omega^2_i - \lambda^2) \right] \cdot \left[ (\tilde{f}^\ddag)^2 + \lambda^2 + \sum_{j}\left(\frac{c^2_j}{\omega^2_j} \frac{\lambda^2}{\omega^2_j - \lambda^2} \right) \right] = 0,
    \end{equation}
    where $c$ is the coupling between the reaction coordinate and each phonon mode and $\omega$ are the phonon frequencies. From Eq.~\ref{eq:barrier_determinant} we see that $\lambda^\ddag$ must satisfy,
    \begin{equation}
        \label{eq:barrier_frequency}
        (\tilde{f}^\ddag)^2 - (\lambda^\ddag)^2 - \sum_{j} \left( \frac{c^2_j}{\omega^2_j} \frac{ (\lambda^\ddag)^2}{\omega^2_j + (\lambda^\ddag)^2} \right) = 0.
    \end{equation}
    Using Eq.~\ref{eq:barrier_frequency} with Eq.~\ref{eq:mem_laplace} clearly shows that $\lambda^\ddag$ satisfies Eq.~\ref{eq:lam_kgh}. 

    In summary, while we can directly apply KGH theory to derive equations for rate constants which depend explicitly on the friction kernel, such equations are a severe simplification of the multidimensional TST formalism we use in the main text.  
    
    \subsection{Desorption rate constants}
\begin{figure*}[thb]  
       \centering
       \includegraphics*[width=3.25in]{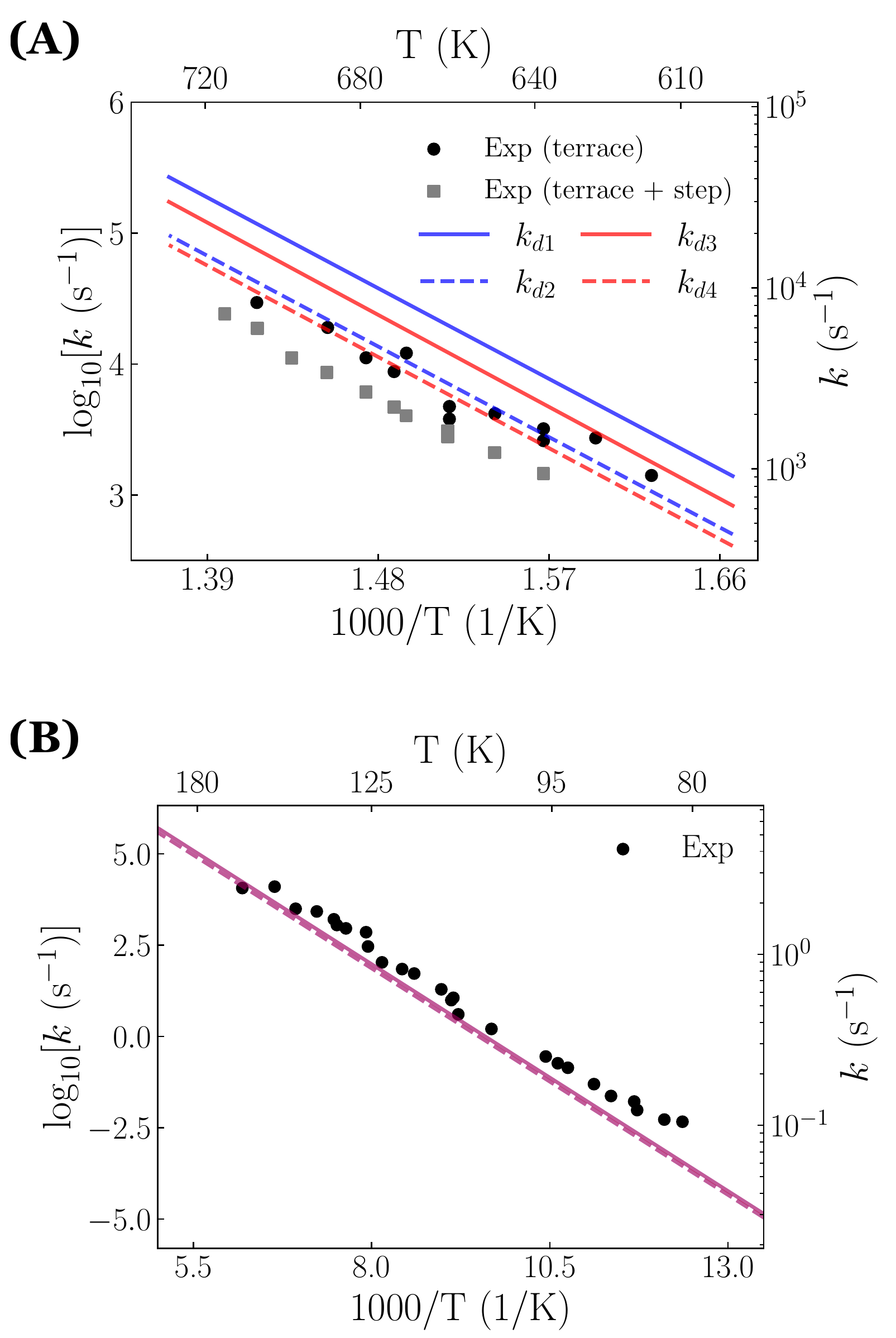}
       \caption{ Comparison of experimental desorption rate constants to theoretical results calculated using a 4x4x8 EMT Pt(111) slab. (A) CO desorption. (B) Xe desorption. }
       \label{figure_s10}
    \end{figure*}

    In the main text, we presented phonon-corrected desorption rates using a model where we computed the friction kernel in the limit of an infinite surface slab by averaging across $k$-space. In Figure~\ref{figure_s10}, we illustrate results from single 4x4x8 surface slab without accounting for phonon dispersion. The results illustrated Figure~\ref{figure_s10} are extremely similar to that of the main text, and the deviation between the two is smaller than the intrinsic uncertainty in the experimental measurements of the rate constant. Parameters used in computing the reaction rates shown in Figure~\ref{figure_s10} are given in Table~\ref{tab:rate_constant_SI}. 

    \begin{table}[htb]
        \caption{Parameters used for computing desorption rate constants shown in Figure~\ref{figure_s10}}
        \label{tab:rate_constant_SI}
        \begin{tabularx}{\textwidth}{XXXXX}
            \toprule
            & $E^{\ddag}$ ($\mathrm{eV}$) & $\omega_{\mathrm{as}}$ ($\mathrm{cm^{-1}}$) & $\tilde{\omega}_{\mathrm{as}}$ ($\mathrm{cm^{-1}}$) & $\tilde{\omega}_{\mathrm{D}}$ ($\mathrm{cm^{-1}}$) \\ 
            \midrule
            CO & 1.47\cite{golibrzuch_co_2015}  & 480\cite{steininger_adsorption_1982} & 133 & 203 \\
            Xe & 0.245\cite{rettner_measurement_1990} & 28\cite{chen_role_2012} & 21 & 156 \\
            \bottomrule
        \end{tabularx}
    \end{table} 
    
    \FloatBarrier
    
\bibliographystyle{unsrt}
\bibliography{main.bib}